\documentclass[aps,prx,reprint,floatfix,showpacs]{revtex4-2}

\usepackage{amssymb}
\usepackage{amsmath}
\usepackage{amsfonts}
\usepackage{graphicx}

\usepackage{cancel}

\newcommand{\fancy}{\mathcal}

\newcommand{\FE}{\kappa}

\newcommand{\wv}{\vec{w}}

\renewcommand{\vec}[1]{\boldsymbol{#1}}
\newcommand{\vech}[1]{\hat{\vec{#1}}}

\newcommand{\Dp}[1]{\partial_{#1}}

\newcommand{\beq}{\begin{eqnarray}}
\newcommand{\eeq}{\end{eqnarray}}

\newcommand{\tr}{\text{Tr}}
\newcommand{\Tr}{{\tr}}

\newcommand{\half}{\tfrac{1}{2}}

\newcommand{\rcite}[1]{Ref.~\onlinecite{#1}}

\newcommand{\Hx}{\text{Hx}}

\newcommand{\xrm}{\text{x}}
\newcommand{\crm}{\text{c}}
\newcommand{\Hrm}{\text{H}}
\newcommand{\xc}{\text{xc}}
\newcommand{\Hxc}{\text{Hxc}}

\newcommand{\epsx}{\epsilon_{\xrm}}
\newcommand{\epsc}{\epsilon_{\crm}}
\newcommand{\epsxc}{\epsilon_{\xc}}
\newcommand{\depsH}{\Delta\epsilon_{\Hrm}}

\newcommand{\RPA}{\text{RPA}}
\newcommand{\QMC}{\text{QMC}}

\newcommand{\rPW}{\text{rPW92}}

\newcommand{\II}[1]{I_{Q\Gamma}\left[#1\right]}
\newcommand{\IB}[1]{\bar{I}(#1)}

\newcommand{\DFA}{{\text{DFA}}}

\newcommand{\SD}{{\text{SD}}}
\newcommand{\DD}{{\text{DD}}}

\newcommand{\fb}{\bar{f}}
\newcommand{\kFb}{\bar{k}_F}
\newcommand{\fmap}{\hat{f}_{\text{x-map}}}

\newcommand{\pr}{^{\prime}}

\renewcommand{\vr}{\vec{r}}

\newcommand{\vrp}{\vec{r}\pr}

\newcommand{\vR}{\vec{R}}

\newcommand{\vq}{\vec{q}}
\newcommand{\vqp}{\vec{q}'}
\newcommand{\vQ}{\vec{Q}}

\newcommand{\iket}[1]{|#1\rangle}

\newcommand{\ibraketop}[3]{\langle#1|#2|#3\rangle}
\newcommand{\ibkouter}[1]{|#1\rangle\langle#1|}

\newcommand{\iout}{\ibkouter}

\newcommand{\bg}{{\text{bg}}}

\newcommand{\LDA}{\text{LDA}}
\newcommand{\LSDA}{\text{LSDA}}
\newcommand{\eLDA}{\text{eLDA}}

\newcommand{\pol}{\text{pol}} 
\newcommand{\unpol}{\text{unpol}} 
\newcommand{\cof}{\text{cofe}} 
\newcommand{\cofe}{\cof} 

\newcommand{\up}{\mathord{\uparrow}}
\newcommand{\down}{\mathord{\downarrow}}

\newcommand{\nh}{\hat{n}}

\renewcommand{\th}{\hat{t}}

\newcommand{\Hh}{\hat{H}}

\newcommand{\Gammah}{\hat{\Gamma}}

\newcommand{\SMSec}[1]{Supp. Mat. Sec.~#1~\cite{SM}}

\usepackage[normalem]{ulem}
\usepackage{color}
\usepackage{xcolor}

\definecolor{Mygrey}{gray}{0.80}
\definecolor{lteal}{rgb}{0.10,0.60,0.70}
\definecolor{dkred}{rgb}{0.50,0.00,0.00}

\newcommand{\comment}[1]{}

\newcommand*{\NoComments}{}%
\newcommand*{\LightComments}{}%

\ifdefined\NoComments
 \newcommand\TG[2]{{#2}}
 
 \newcommand\Look[1]{\textcolor{dkred}{#1}}
\else
 \ifdefined\LightComments
  \newcommand\TG[2]{\textcolor{blue}{#2}}
  
  \newcommand\Look[1]{\textcolor{dkred}{#1}}
 \else
  \newcommand\TG[2]{\textcolor{blue}{ \sout{#1}{#2}}}
  
  \newcommand\Look[1]{\textcolor{dkred}{#1}
 \fi
\fi

\begin{document}

\title{Local density approximation for excited states}

\author{Tim Gould}
\affiliation{Queensland Micro- and Nanotechnology Centre, %
  Griffith University, Nathan, Qld 4111, Australia}
\email{t.gould@griffith.edu.au}
\author{Stefano Pittalis}
\affiliation{CNR-Istituto Nanoscienze, Via Campi 213A, I-41125 Modena, Italy}

\begin{abstract}
The ground state of an homogeneous electron gas is a paradigmatic state that has been used to model and predict the electronic structure of matter at equilibrium for nearly a century.
For half a century, it has been successfully used to predict ground states of quantum systems via the local density approximation (LDA) of density functional theory (DFT); and systematic improvements in the form of generalized gradient approximations and evolution thereon.
Here, we introduce the LDA for \emph{excited} states by considering a particular class of non-thermal ensemble states of the homogeneous electron gas.
These states find sound foundation and application in ensemble-DFT -- a generalization of DFT that can deal with ground and excited states on equal footing.
The ensemble-LDA is shown to successfully predict difficult low-lying excitations in atoms and molecules for which approximations based on local spin density approximation (LSDA) and time-dependent-LDA fail.
\end{abstract}
 
\maketitle


\section{Introduction}

Excitation of many-electron systems characterize novel states of matter and increasingly permeate the functions of novel advanced technologies. 
\TG{}{In problems ranging from photovoltaic devices to quantum dots, nano-particle catalysts, and quantum computing devices --- particle-like, collective, or topological excitations are exploited coherently.}
Challenges are multidisciplinary, yet solutions can be inspired -- and, increasingly, predicted -- by computationally investigating quantum structures and mechanisms at the nanoscale.
Density functional theory~\cite{HohenbergKohn,KohnSham} (DFT) has dominated the stage of computational electronic structure methodologies since the 1960s, by balancing accuracy with efficiency.
But DFT does not handle excited states directly, being restricted to addressing eigenstates of lowest energy (i.e. ground states). 
This work will show how successful DFT methods for ground states can be upgraded into methods for {\em also} tackling excited states.

The most fundamental model from which DFT gained inspiration, can be traced back to the seminal works by Thomas and Fermi~\cite{Thomas1927,Fermi1927}.
In 1927, they independently proposed a remarkable approximation for quantum physics -- that the state of any many-electron system can be modelled by referring, via the particle density (a local quantity), to an 
homogenous gas of electrons. 
Due to its poor treatment of kinetic energy contributions, the resulting Thomas-Fermi approximation is not very good in practice.
But almost all modern modelling of electronic structure employs its spiritual descendent, in the form of Kohn-Sham DFT~\cite{HohenbergKohn,KohnSham}:
1) kinetic energy contributions are treated quantum mechanically, via a non-interacting auxilliary system;
2) the energy of electrostatic interactions is treated clasically, for any given particle density;
3) the HEG is \emph{only} used to treat the remaining quantum exchange-correlation (xc) energy contributions.

The homogeneous electron gas (HEG~\footnote{AKA jellium or uniform electron gas -- we shall use HEG exclusively throughout}) is, arguably, the simplest many-electron system.
It involves {$N\to\infty$} electrons interacting in response to a uniform positive background charge of fixed density, $n$, and volume, $V=N/n\to\infty$.
The resulting (interacting) electronic structure problem can be solved semi-analytically in its high-density and low-density limits, and to high accuracy for moderate densities using quantum Monte Carlo (QMC) techniques.~\cite{Ceperley1980,Zong2002,Spink2013}
The \emph{known} paradigmatic xc behaviour of HEG may then be used to approximate the \emph{unknown} xc behaviour of inhomoheneous quantum systems, via parametrisations.~\cite{VWN,PW92,Chachiyo}.
Crucially, it has also been recognised that the LDA provides exact leading terms in a semi-classical expansion of any quantum system, under appropriate limits;~\cite{Lieb1973,Lieb1977,Lieb1981,Elliott2008,Burke2016}
which helps to explain the ongoing success of the Jacob's ladder~\cite{Perdew2001-Jacob} philosophy of systematically improving on the LDA.~\cite{Becke1988,PBE,SCAN,wB97x}

{\em What about excited states?}
In the late 1980s, the time-depended extension of DFT (TDDFT) was revealed to be an highly effective tool for simulating spectra, via a perturbative (linear-response) expansion around the ground state.
But, despite its ongoing success, it was soon revealed~\cite{Maitra2004-Double,Maitra2005} that approximations to TDDFT could not describe important double excitations at all; and struggle to describe charge transfer excitations except by using specialized approximations.~\cite{Iikura2001,Stein2009,Maitra2017-CT}
More recently, singlet-triplet inversion~\cite{deSilva2019} (with great promise for photovoltaics) has emerged as another important problem where TDDFT struggles.~\cite{Ehrmaier2019,Ghosh2022}

In parallel with TDDFT, Kohn and collaborators put forward a density functional theory for {\em stationary} excitations based on mixed states (ensembles) rather than pure states: ensemble-DFT (EDFT).~\cite{GOK-1,GOK-2} 
Unlike the perturbation-based formalism of TDDFT, EDFT recast the problem of computing excited states into an extended ``ground state''-like problem involving variational minima.
TDDFT's rapid success in predicting spectra, and challenges in constructing useful ensemble approximations, initially led to EDFT falling by the wayside.
Recently, however, it has re-emerged as a powerful alternative to TDDFT because approximations in EDFT can solve precisely those excitation problems for which TDDFT struggles or fails.~\cite{Pastorczak2014-GI,Filatov2015-Double,Filatov2016,Pribram-Jones2014,Yang2014,Yang2017-EDFT,Gould2018-CT,Sagredo2018,Fromager2019-single,Loos2020-EDFA,Marut2020,Gould2020-Molecules,Gould2021-DoubleX,Gould2022-pEDFT}

Moreover, recent theoretical breakthroughs~\cite{Yang2017-EDFT,Gould2017-Limits,Gould2019-DD,Fromager2020-DD,Gould2020-FDT,Gould2023-ESCE} have revealed aspects of the architecture of key functional forms in EDFT that have opened unprecedented possibilities for novel approximations for excited states.
The change of perspective brought about by EDFT compared to (TD)DFT is radical: 
1) the auxiliary states of the Kohn-Sham ensemble can acquire the form of coherent (finite) superposition of Slater determinants (rather than the  `disentangled' single determinant for pure ground states);
2) the ensemble Hartree energy (in contrast to the {\em classical} Hartree energy) accounts for peculiar quantum features;
3) the ensemble exchange energy does not (necessarily) reduce to textbook Fock-exchange expressions;
4) in addition to regular-looking state-driven correlations, unusual density-driven correlations emerge.

In this work, we demonstrate that the same system of knowledge allows us to derive an exchange-correlation energy approximation from first principles (\emph{ab initio}).
We consider the prominent example of approximations that are derivable from the HEG.
Given nearly 100 years of exploration, one might expect the HEG to have given up all its useful secrets.
Crucially this work reveals that when the HEG is viewed from the perspective of EDFT, we can introduce a class of non-thermal ensembles from which we can derive a local approximation for excited states {\em directly}. 
The regular LDA has provided an highly-effective cornerstone for systematic improvements for ground states -- both as the first rung of Jacob's ladder~\cite{Perdew2001-Jacob} and as a paradigmatic/semi-classical limit that can constrain functional forms~\cite{Becke1988,PBE,SCAN,DM21}.
The ensemble-LDA developed in this work therefore provides us with a (long-sought) cornerstone for systematic improvements to approximations for excited states.

The remainder of this work is organized as follows:
Section~\ref{sec:Theory} gives an introduction to the HEG in the context of density functional theory \TG{}{and briefly reviews relevant attempts that pre-date our current proposal.
Section~\ref{sec:EDFTcof} introduce the elements of ensemble-DFT which are exploited in the novel parts of the work and presents the relevant ensemble-states of HEG, which are designed to capture excited-state physics in crucial energy components of the HEG ensemble-states (Appendix~\ref{app:Param} reports a parametrisation).
Sections~\ref{sec:Inhomogeneous} and \ref{sec:Applications} demonstrate the practical usefulness of the formal developments done by setting up (Sec.~\ref{sec:Inhomogeneous}) and applying (Sec.~\ref{sec:Applications}) an ensemble-LDA to atoms and molecules.}
Finally, Section~\ref{sec:Conclusions} summarizes the work, looks toward the near future, and draws conclusions.

\section{Local density approximation}
\label{sec:Theory}

\TG{}{
This section will first motivate the standard approach to understanding HEGs, in the context of density functional theory, to lay out the foundation of the local density approximation (LDA).
The next Section~\ref{sec:EDFTcof} then introduces ensemble density functional theory and excited state HEGs as the foundations of the excited state LDA (eLDA).
Together, these sections provide the key theoretical tools for the rest of the work.}
Throughtout, we use atomic units so that lengths are expressed in Bohr and energies are expressed in Hartree (Ha).

Before getting to details, it is worth noting that the properties of HEGs are conventionally defined using the the Wigner-Seitz radius, $r_s:=(\tfrac{3}{4\pi n})^{1/3}\approx 0.620350 n^{-1/3}$, and spin-polarization factor, $\zeta=\tfrac{n_{\up}-n_{\down}}{n}$. 
Here, $n$ is the density of electrons and $n_{\up,\down}$ are the densities of $\up,\down$ electrons obeying $n_{\up}+n_{\down}=n$.
This combination of terms reflects the fact that interactions between same- and different-spin electrons are fundamentally different due to the Pauli exclusion principle, so energies change not only with the total density but also the relative contributions of majority ($\up$) and minority ($\down$)  lectrons to the density.
\TG{}{We will sometimes use $n$ instead of $r_s$, to clarify dependence on the density.}

\subsection{DFT of homogeneous electron gases}

Density functional theory (DFT) provides an important
tool for the analysis and parametrisation of HEGs.
Key theorems~\cite{HohenbergKohn,Levy1979,Lieb1983}
demonstrate that all properties of a quantum mechanical
ground state are described by its density, $n(\vr)$
(constant, $n$ in an HEG). This is easily extended to
spin-DFT,~\cite{vonBarth1972} which covers \emph{de facto}
ground states like the lowest energy with a given
spin-polarization, $\zeta(\vr)$ (constant $\zeta$ in an HEG).
DFT is typically used synonymously with Kohn-Sham (KS)
DFT,~\cite{KohnSham} and we shall adopt this
convention throughout.

In Kohn-Sham DFT, the ground state
energy of an $N$-electron
system in external (nuclear) potential, $v(\vr)$,
is written as,
\begin{align}
E_0[n]:=&T_s[n] + \int n v d\vr + E_{\Hrm}[n]
+ E_{\xrm}[n]  + E_{\crm}[n]\;,
\label{eqn:E0}
\end{align}
where $[n]$ indicates a functional of the density,
$n(\vr)$, obeying $\int n d\vr=N$.
Useful exact energy expressions are known for:~\footnote{Note, here we mean `exact' in the sense that of the exact form, not the the functional obtained at the exact density and orbitals.}
\begin{enumerate}
\item The Kohn-Sham kinetic energy functional, $T_s[n]$,
that includes kinetic energy effects from a \emph{non-interacting}
system with the same density (and spin)
-- we may write $T_s=\sum_{i\sigma\in\text{occ}}
\int\half|\nabla\phi_{i\sigma}(\vr)|^2d\vr$
using a set of occupied Kohn-Sham orbitals,
$\phi_{i\sigma}(\vr)$;~\cite{KohnSham}
\item The Hartree energy functional,
$E_{\Hrm}[n]=U[n]$,
that includes mean-field electrostatic interactions;
\item The Fock exchange energy functional,
$E_{\xrm}[n]=-\sum_{ii' \sigma\in\text{occ}}
U[\phi_{i\sigma}\phi_{i'\sigma}^*]$,
that includes corrections for Fermionic exchange based
on the same non-interacting system used for $T_s$.
\end{enumerate}
The \emph{unknown} correlation energy functional,
$E_{\crm}[n]$, captures classical and quantum
contributions that are missed in the other terms.

Here we introduced an electrostatic Coulomb integral,
\begin{align}
U[\rho]=\int \rho(\vr)\rho^*(\vrp)\frac{d\vr d\vr'}{2|\vr-\vr'|}
=U[\rho^*]
\label{eqn:Uee}
\end{align}
that was adapted for complex-valued inputs to accommodate
classical (here, in $E_{\Hrm}$ only) and quantum
(here, in $E_{\xrm}$ only) interactions.
All functionals are readily extended to spin-polarized
ground states by introducing the number, $N_{\up}\leq N$,
of $\up$ electrons ($N_{\down}=N - N_{\up}$) as an
additional constraint, or equivalently setting
$\zeta=\tfrac{N_{\up}-N_{\down}}{N}$.
Precise details do not matter at this point and will be
introduced as required.

In a standard HEG, the mean-field Hartree contribution
(from $E_{\Hrm}$) is cancelled exactly by the
positive background charge.
The energy per particle, $e=E/N$, of an HEG may
therefore be separated into three components,
\begin{align}
e(n,\zeta)=t_s(n,\zeta)+\epsx(n,\zeta)+\epsc(n,\zeta)\;,
\label{eqn:eHEG}
\end{align}
using eq.~\eqref{eqn:E0}. Here, $n$ and $\zeta$ are scalar constants; 
and $t_s:=T_s/N$, $\epsx:=E_{\xrm}/N$ and $\epsc:=E_{\crm}/N$
are energy densities per particle.
The Kohn-Sham kinetic and exchange energies may
be obtained analytically, and are,
\begin{align}
t_s(r_s,\zeta)=&t_s(r_s)
\tfrac{(1+\zeta)^{5/3} + (1-\zeta)^{5/3}}{2}
:=t_s(r_s)f_s(\zeta)\;,
\label{eqn:tszeta}
\\
\epsx(r_s,\zeta)=&
\epsx(r_s)
\tfrac{(1+\zeta)^{4/3} + (1-\zeta)^{4/3}}{2}
:=\epsx(r_s)f_{\xrm}(\zeta)\;,
\label{eqn:epsxzeta}
\end{align}
where,
\begin{align}
t_s(r_s):=&\tfrac{C_t}{r_s^2}=\tfrac{3}{10}\big( \tfrac{9\pi}{4} \big)^{2/3}r_s^{-2}
=1.10495r_s^{-2}\;,
\label{eqn:ts}
\\
\epsx(r_s):=&\tfrac{-C_x}{r_s}=-\tfrac{3}{4\pi}\big( \tfrac{9\pi}{4} \big)^{1/3}r_s^{-1}
=-0.458165r_s^{-1}\;,
\label{eqn:epsx}
\end{align}
are the kinetic and exchange energies of an unpolarized
HEG (in atomic units).
We may alternately write,
$t_s(n)=2.87123n^{2/3}$
and $\epsx(n)=-0.738559n^{1/3}$.

The final ingredient is the correlation energy term,
\begin{align}
\epsc(r_s,\zeta):=\sum_k \epsc^k(r_s)f_{\crm}^k(\zeta)\;,
\label{eqn:epsczeta}
\end{align}
which has known series expansions for the
high- ($r_s\to 0$) and low-density ($r_s\to\infty$) limits,
but is unknown in general.
Total energies, $e^{\QMC}$, of HEGs may be evaluated
to high accuracy via quantum Monte-Carlo (QMC)
simulations, which have served to supplement limiting
cases since pioneering
work by Ceperley and Alder.~\cite{Ceperley1980}
Then, $\epsc=e^{\QMC}-t_s-\epsx$, may be
parametrised (e.g.~\cite{VWN,PW92,Chachiyo})
by a truncated series in the general form
of \eqref{eqn:epsczeta}. Models and parameters for $\epsc$
are usually designed to satisfy or approximately satisfy
limiting behaviours of HEGs, with some free
parameters that can be optimized to reproduce
reference data from QMC at intermediate values.

\begin{table*}
\caption{Summary of Kohn-Sham derived properties
of the HEGs considered in this work.
Here, $C_s=1.10495$ and $C_{\xrm}=0.458165$.
The cases $\zeta=0$ and $\fb=2$ correspond to an unpolarized gas;
and $\zeta=1$ and $\fb=1$ are equivalent.
\label{tab:HEGs}}
\begin{ruledtabular}\begin{tabular}{lccccc}
Type of HEG & Params & $t_s$ & $\epsx$ & $\depsH$
\\\hline
Unpolarized gas & $r_s$
& $\tfrac{C_s}{r_s^2}$
& $\tfrac{-C_{\xrm}}{r_s}$
& 0
\\
Polarized gas & $r_s$, $\zeta$
& $\tfrac{C_s}{r_s^2}\tfrac{(1+\zeta)^{5/3}+(1-\zeta)^{5/3}}{2}$
& $\tfrac{-C_{\xrm}}{r_s}\tfrac{(1+\zeta)^{4/3}+(1-\zeta)^{4/3}}{2}$
& 0
\\
Constant occupation factor (cof) & $r_s$, $\fb$
& $\tfrac{C_s}{r_s^2}\big[\tfrac{2}{\fb}\big]^{2/3}$
& $\tfrac{-C_{\xrm}}{r_s}\big[\tfrac{2}{\fb}\big]^{1/3}$
& $|\epsx|\tfrac{(2-\fb)(\fb-1)}{\fb}$
\end{tabular}\end{ruledtabular}
\end{table*}

\subsection{From HEGs to real systems}
\label{sec:LDA}

\TG{}{
The local density approximation (LDA) models the quantum mechanics of an inhomogeneous system by combining exact DFT terms with terms reusing expressions from HEGs.
In detail, for an inhomogeneous system with nuclear potential, $v$, and electronic density $n$, the LDA yields a ground state energy,
\begin{align}
E_{\LDA}[n]=&T_s[n] + \int n v d\vr + E_{\Hrm}[n] + \int n \epsxc d\vr\;.
\label{eqn:LDA}
\end{align}
The first three terms of \eqref{eqn:LDA} have the same form as the corresponding exact terms. 
HEGs are used to approximate the xc energy, $E_{\xc}[n] \approx E^\LDA_{\xrm}[n]+E^\LDA_{\crm}[n]$
Specifically, the xc energy density is locally approximated by using the xc energy per unit particle, $\epsxc(\vr)=\epsx(n(\vr))+\epsc(n(\vr))$, [from Eqs.~\eqref{eqn:epsx} and \eqref{eqn:epsczeta}] of an HEG with density $n(\vr)$.
}

\TG{}{
The LDA is an important component of many successful DFAs and is the cornerstone of ``Jacob's ladder''~\cite{Perdew2001-Jacob} for DFAs.
Its success is typically justified by arguments based on `xc holes' that also help in extending LDAs to incorporate \emph{semi-}local properties like density gradients or meta-densities.~\cite{PribramJones2015}
Successes of the Thomas-Fermi approximation (TFA) -- which also approximate the non-interacting kinetic energy by an LDA -- and of the usual KS LDA have also been justified by appealing to exact semi-classical limits and universal bounds for quantum mechanical systems.~\cite{Lieb1973,Lieb1977,Lieb1981,LiebOxford,Elliott2008,Burke2016}
It should not be forgotten, however, that without the initial judicious choice by Kohn and Sham of {\em not} changing the form 
of $T_s$ and $E_{\Hrm}$, the broad success of DFT would not have gone much beyond the much more limited success of the TFA.
}

\TG{}{
One of the most important and earliest extensions of the LDA is 
the local \emph{spin} density approximation (LSDA),~\cite{Gunnarsson1976}
which uses the local densities, $n_{\up}(\vr)$ and $n_{\down}(\vr)$, of the two spin-channels separately.
Eq.~\eqref{eqn:LDA} is extended to, $E_{\LSDA}[n_{\up},n_{\down}]=T_s[n_{\up},n_{\down}] + \int n v d\vr + E_{\Hrm}[n] + \int n \epsxc(n,\zeta) d\vr$,
which includes the effect of spin in $T_s$ by allowing the orbitals to differ for different spin channel and uses the local density, $n(\vr)=n_{\up}(\vr)+n_{\down}(\vr)$, and spin-polarization, $\zeta(\vr)=\tfrac{n_{\up}(\vr)-n_{\down}(\vr)}{n(\vr)}$, in the HEG parametrisation.
The LDA then becomes the special case $n_{\up}=n_{\down}$.
Indeed the LSDA is the true cornerstone of almost all modern DFT applications because, by introducing spin, it extends the applicability of HEG-based approximations to spin-polarized states: i.e., either ground states acted upon a (collinear) magnetic field, or the lowest excited state  of a given net spin-polarization.
A natural next step would be to extend the LDA to stationary but otherwise general \emph{excited} states.
}

\subsection{A long standing conundrum: local approximations for excitations}
\label{eqn:ELDAEarly}

\TG{}{
Let us begin by highlighting why the excited state LDA (eLDA) problem is more difficult than ground states.
Firstly, DFT itself is not well-defined for excited states so one needs to work out what terms should be treated exactly, or by approximations, or other extensions.~\cite{CWV08,Levi2020}
Secondly, the spectra of molecules is discrete and the spectra of insulators is gapped, whereas HEGs are a metal with no gap and a continuous (dense) spectrum; the eLDA must therefore map between fundamentally different physics.
Finally, excitations are in some sense non-local (because variation in the single-particle orbitals must be sharply described), which challenges the use of only local properties like the particle-density and local spin-polarisation.
Despite these fundamental difficulties, various attempts have been made to produce eLDAs, as surveyed below.}

\TG{}{An early attempt was by Kohn,~\cite{Kohn1986}, who sought to connect inhomogeneous excited states with finite temperature HEGs by enforcing a relation between auxiliary ensembles of excited states and proper thermal ensembles.
Kohn's approach implies to deal with quantities like the entropy, heat capacity, and temperature thus reducing the problem of finding excitation energies to the problem of finding some key aspect of the thermodynamics of a system.
Because the approach involves an effective temperature to be estimated by integrating entropies per unit volume over the {\em whole} space [see Eq. (26) in Ref. \cite{GOK-3} for a helpful discussion], Kohn does not regard the approach as a purely local-density approximation but defines it to be a {\em quasi}-LDA.
This approximation was used in a foundational paper on GOK-EDFT,~\cite{GOK-3} but its restriction to averages over multiple excited states meant that it could not resolve singlet-triplet spin splitting.}

\TG{}{Theophilou and Papaconstantinou~\cite{Theophilou2000} later introduced an eLDA in which the reference to thermodynamics was finally removed.
They also added the important spin dependence, useful to evaluate spin splittings.
Their approach reduces to the LSDA in which the spin polarization is replaced with a global quantity (independent of space) that is related to the spin state of the underlying  spin-restricted symmetry adapted approach, and is therefore a quasi-LSDA.}

\TG{}{Harbola and co-workers~\cite{Samal2005,Hemanadhan2014} were able to exploit the exchange energy of {\em proper} excited HEGs to derive an LDA thereto.
Their approach is most similar in spirit to the work presented here.
However, their approach does not go beyond exchange, possibly because they used the conventional ground state DFT framework instead of EDFT.
Furthermore, orbital-dependent self-interaction-correction terms must be included as well.}

\TG{}{A much more recent attempt, that works within the framework of EDFT, is from Loos, Fromager and coworkers~\cite{Loos2020-EDFA,Marut2020} who parameterized a local density approximation for ensembles based on the properties of uniform electron gases with \emph{finite numbers of electrons}.~\cite{Loos2012,Loos2013-Ringium,Loos2014-Ringium,Loos2014}
Their approach captures important excited-state physics, notably by avoiding the issue of a continuous spectra in the HEG, and is explicitly designed for EDFT problems.
However, applications were presented only for two-electron systems, and extensions are required to make the approach more generally applicable.}

\TG{}{Finally, ``combination rules'' have recently been identified that allow re-use of \emph{any} existing ground state DFA, including L(S)DA, for use in excited state problems.~\cite{Gould2021-DoubleX,Gould2022-pEDFT}
Therefore, combination rules highlight that the locality assumptions behind LDAs \emph{can be extended to excited states}.
This is because combination rules are equivalent to setting (for a specific excited state) $n\epsxc^{\text{excited}}(n) \approx \sum_P c_P n_P\epsxc^{\DFA}(n_P)$ in $E_{\xc}^{\text{excited}}=\int n\epsxc^{\text{excited} }d\vr$, where $P$ labels auxilliary states, $n_P$ are their densities, $c_P$ are constants and $\epsxc^{\DFA}$ is an existing (semi-)local DFA.
It follows that the locality of $\epsxc^{\DFA}$ is extended through the weighted sum in $\epsxc^{\text{excited}}$.
Despite being exact for exchange and working effectively for xc DFAs,~\cite{Gould2021-DoubleX} the rules are \emph{inexact} for correlations meaning there is room for improvement.
The approach presented here may thus be regarded as a first step toward circumventing combination rules, by replacing them with an excited state LDA foundation.}

\TG{}{Crucially, we should stress that previous attempts (except the combination rules)  were made by working with an {\em incomplete} understanding of the structure of the relevant exact density functionals for excited states.
In particular,  recent analysis has revealed that Hartree and exchange in EDFT go beyond the previous restrictive --  sometimes even problematic -- definitions.~\cite{Gould2017-Limits,Gould2020-FDT}
We now know that correlations come in two kinds: state-driven (which resemble ground-state correlations) and density-driven (which are totally new).
All new terms in the ensemble Hartree and DD correlation are highly non-local expression.
Recent progress has also revealed that regular DFT approximations (and thus LDA expressions) are appropriate for the novel exchange terms and the novel SD-correlation terms~\cite{Gould2020-FDT,Gould2021-DoubleX} -- more below.
Thus, next in Sec.~\ref{sec:EDFTcof}, we show that the aforementioned novel components can be determined {\em also} in HEGs.
}

\section{EDFT of excited state HEGs}
\label{sec:EDFTcof}

\TG{}{
As discussed in Sec.~\ref{eqn:ELDAEarly} previous attempts to develop excited state LDAs have run into problems or limitations.
It is reasonable to assume that some of these difficulties reflect the fact that previous work was based on incomplete understanding of the structure of excited states.
This section thus first (Sec.~\ref{sec:EDFT}) discusses an upgraded and first principles understanding of how the excited states physics get encoded into
the relevant energy components of EDFT.
From these foundations it becomes easier to discern which HEG expressions may be used to approximate which exact energy component in inhomogeneous systems, as a first enabling step toward an effective excited state LDA.%
}

\TG{}{
Section~\ref{sec:cof} then introduces and derives a `constant occupation factor ensemble' (cofe) HEG to serve as a foundation for excited state approximations.
The results derived in this section 
are later applied to inhomogeneous systems in Sec.~\ref{sec:Inhomogeneous}, which also expands on why/how the cofe HEG is relevant.
Results on realistic systems are then presented in Sec.~\ref{sec:Applications}.}

\subsection{Ensemble DFT from first principles}
\label{sec:EDFT}

To understand ensemble DFT,~\cite{GOK-1,GOK-2} let us first define quantum state ensembles.
A (quantum state) ensemble, $\Gammah$, is an operator that describes a classical mixture of quantum states.
It may be defined using a spectral representation,
\begin{align}
\Gammah=&\sum_{\FE}w_{\FE}\iout{\FE}\;,~~~
0\leq w_{\FE}\leq 1,~~~\sum_{\FE}w_{\FE}=1\;,
\end{align}
in which an arbitrary set of orthonormal
quantum states, $\iket{\FE}$,
are assigned probabilities/weights, $w_{\FE}$.
Operator expectation values,
$\bar{O}=\ibraketop{\Psi}{\hat{O}}{\Psi}$,
are replaced by $\bar{O}^{\wv}=\tr[\Gammah^{\wv}\hat{O}]
=\sum_{\FE}w_{\FE}\ibraketop{\FE}{\hat{O}}{\FE}$
which involves quantum and classical averages.
Ensembles are more flexible than wave functions,
so can describe constrained, open and
degenerate systems that are otherwise outside the
remit of wave function mechanics or DFT.
Various theorems~\cite{Perdew1982,GOK-1,GOK-2}
extend key results of DFT to ensembles, including
important variational principles.

In excited state EDFT, the usual variational formula,
$E_0=\min_{\Psi}\ibraketop{\Psi}{\Hh}{\Psi}$,
is replaced by the weighted average,
\begin{align}
E^{\wv}:=& \inf_{\Gammah^{\wv}}\Tr\big[\Gammah^{\wv}\Hh\big]
=\sum_{\FE}w_{\FE}E_{\FE}\;,
\label{eqn:Ew}
\end{align}
where $\Gammah^{\wv}$ is an ensemble with a given
set of weights $\wv=\{w_0,w_1,\ldots\}$;
and $E_{\FE}$ are eigen-energies of $\Hh$ ordered
such that the lowest energies are associated with the
largest weights. The energies are in usual ascending
`excitation' order if we define the weights to be
monotonically decreasing, i.e. $w_{\FE'}\leq w_{\FE}$ for
$E_{\FE'}\geq E_{\FE}$.
Note, we follow the usual convention of using superscripts $^{\wv}$ (or $^{\cof}$ later) to identify ensemble functionals.
But we depart from the recent convention of using calligraphic letters to avoid confusion between
$\fancy{E}$ for total energies of ensembles, and $\epsilon$ for energies per particle of HEGs.

It is convenient to generalize eq.~\eqref{eqn:E0} to ensembles by writing,
\begin{align}
E^{\wv}[n]:=&T_s^{\wv}[n] + \int n v d\vr + E_{\Hrm}^{\wv}[n]
+ E_{\xrm}^{\wv}[n] + E_{\crm}^{\wv}[n]\;.
\label{eqn:E0w}
\end{align}
Here, $\wv$ indicates the set of weights, $n$ is the density, and $v$ is the external potential.
In a Kohn-Sham formalism, the ensemble density is conveniently written as,
\begin{align}
n^{\wv}(\vr):=&\sum_i f_i^{\wv}n_i(\vr)\;,
&
f_i^{\wv}:=&\sum_{\FE}w_{\FE}\theta_i^{\FE}
\;,
\label{eqn:nw}
\end{align}
in terms of orbital densities, $n_i(\vr):=|\phi_i(\vr)|^2$; and average occupation factors, $f_i^{\wv}$, which may be non-integer and involve a weighted average over
the integer occupation factors, $\theta_i^{\FE}\in(0,1,2)$ (i.e. no occupation, occupation in one spin, or occupation in both spins) of each KS state in the ensemble.
The orbitals obey a spin-independent KS-like equation, $[\th+v_s^{\wv}(\vr)]\phi_i(\vr)=\varepsilon_i\phi_i(\vr)$, where $\th\equiv-\half\nabla^2$ is the one-body kinetic
energy operator.
\TG{}{Note, functions of the position like orbitals ($\phi_i\equiv \phi_i^{\wv}$) and densities ($n_i\equiv |\phi_i^{\wv}|^2$ and similar) also carry  an {\em implicit} dependence on the weights, $\wv$, as do KS wave functions ($\iket{\FE_s}\equiv \iket{\FE_s^{\wv}}$), but we leave the superscript off to avoid clutter in equations.}

With the ensemble formalism defined, we are now ready
to define the terms in eq.~\eqref{eqn:E0w}.
Recent work~\cite{Gould2017-Limits,Gould2019-DD,Gould2020-FDT}
has sought to rigorously define exact energy
functionals for excited state ensembles, giving,
\begin{align}
T_s^{\wv}[n]:=&\sum_i f_i^{\wv}\int\half|\nabla\phi_i|^2d\vr\;,
\label{eqn:Tsw}
\\
E_{\Hrm}^{\wv}[n]:=&\sum_{\FE\FE'} w_{\max(\FE,\FE')}U[n_{s,\FE\FE'}]\;,
\label{eqn:EHw}
\\
E_{\xrm}^{\wv}[n]:=&-\sum_{ii'} f^{\wv}_{\max(i,i')} U[\phi_i\phi_{i'}^*]
\label{eqn:Exw}
\end{align}
Here, we used $\int \phi^*\th\phi d\vr=\int \half|\nabla\phi|^2d\vr$; $U[\rho]$ as defined earlier in Eq.~\eqref{eqn:Uee}; introduced $n_{s,\FE\FE}(\vr)=\ibraketop{\FE_s}{\nh(\vr)}{\FE_s}$ as the density of Kohn-Sham state, $\iket{\FE_s}$; and introduced $n_{s,\FE\neq\FE'}(\vr)=\ibraketop{\FE_s}{\nh(\vr)}{\FE'_s}$ as the (potentially complex-valued) transition density between Kohn-Sham states $\iket{\FE_s}$ and $\iket{\FE'_s}$.
\TG{}{Eq.~\eqref{eqn:Tsw} retains its well-known ``textbook'' expression in all cases.
By contrast, eqs.~\eqref{eqn:EHw} and \eqref{eqn:Exw} \emph{reproduce} textbook expressions for the lowest energy state of each spin-polarization (maximal $|S_z|$) only -- i.e `conventional' states accessible by ground state DFT -- but are different in ensembles and excited states.}

The remaining energy, $E_{\crm}^{\wv}:=E^{\wv} - \int n v d\vr  - T_s^{\wv}  - E_{\Hrm}^{\wv} - E_{\xrm}^{\wv}$, is the unknown correlation energy functional.
It is convenient to partition,
\begin{align}
E_{\crm}^{\wv}[n]:=&
E_{\crm}^{\SD,\wv}[n] + E_{\crm}^{\DD,\wv}[n]\;.
\label{eqn:Ecw}
\end{align}
into state-driven (SD) and density-driven (DD) components, each with different physical origins.~\cite{Gould2019-DD,Fromager2020-DD,Gould2020-FDT}
\TG{}{The SD correlation energy may be written as,
\begin{align}
E_{\crm}^{\SD,\wv}:=&\int_0^1d\lambda
\int_0^{\infty}\frac{-d\omega}{\pi}
\int \frac{d\vr d\vrp}{2|\vr-\vrp|}
\nonumber\\&\times
\big[\chi_{\lambda}^{\wv}(\vr,\vrp;i\omega)-\chi_0^{\wv}(\vr,\vrp;i\omega)\big]\;,
\label{eqn:EcSDw}
\end{align}
in terms of density-density response function [see \rcite{Gould2020-FDT} for details], and retains the same general expression as the ground state correlation energy.
The density-driven term is always zero in conventional states (e.g. molecular ground states) so is not considered in ground state DFT.
We shall address its general expression shortly.
}

It is sometimes useful to rewrite eqs~\eqref{eqn:EHw}
and \eqref{eqn:Exw} as,
\begin{align}
E_{\Hrm/\xrm}^{\wv}=&\int n_{2,\Hrm/\xrm}^{\wv}(\vr,\vrp) \frac{d\vr d\vrp}{2|\vr-\vrp|}
\label{eqn:Ewn2}
\end{align}
using the ensemble Hartree and exchange pair-densities,
\begin{align}
n_{2,\Hrm}^{\wv}(\vr,\vrp)=&\sum_{\FE\FE'}w_{\max(\FE,\FE')}
n_{s,\FE\FE'}(\vr)n_{s,\FE'\FE}(\vrp)\;,
\label{eqn:n2H}
\\
n_{2,\xrm}^{\wv}(\vr,\vrp)=&-\sum_{ii'}f_{\max(i,i')}
\rho_i(\vr,\vrp)\rho_{i'}^*(\vr,\vrp)\;,
\label{eqn:n2x}
\end{align}
where $\rho_i(\vr,\vrp)=\phi_i(\vr)\phi^*_i(\vrp)$.
It is straightforward to see that using eqs~\eqref{eqn:n2H} and \eqref{eqn:n2x} in \eqref{eqn:Ewn2} give the same energies as
\eqref{eqn:EHw} and \eqref{eqn:Exw}, respectively.
\TG{}{The DD correlation energy also can be expressed using \eqref{eqn:Ewn2}.
Its pair-density has a similar form to the Hartree energy,
\begin{align}
n_{2,\crm}^{\DD,\wv}(\vr,\vrp)=&\sum_{\FE\FE'}w_{\max(\FE,\FE')}
\int_0^1 
\big[
n^{\lambda}_{\FE\FE'}(\vr)n^{\lambda}_{\FE'\FE}(\vrp)
\nonumber\\&
- n_{s,\FE\FE'}(\vr)n_{s,\FE'\FE}(\vrp) \big]d\lambda\;,
\label{eqn:n2cDD}
\end{align}
but involves the difference between transition densities at interaction strength $\lambda$ and their KS counterparts (i.e. $\lambda=0$).~\cite{Gould2020-FDT}
Thus, like the Hartree energy, $E_{\crm}^{\DD,\wv}$ has an explicitly non-local dependence on densities and orbitals and should not by approximated locally.}
Details and other helpful relationships for functionals will be introduced and used as required.

Before proceeding further, we make the important assumption that the results of Section~\ref{sec:EDFT} apply to HEGs.
This is an assumption because all EDFT results shown so far are for \emph{finite} systems with
\emph{countable} numbers of excitations.
By contrast, homogeneous electron gases are \emph{infinite} and their excitations are
\emph{uncountable}.
The rest of this manuscript treats HEGs as the appropriate thermodynamic limit of finite
systems whose properties are consistent with the ensemble density functional theory presented
in this section, and so obey straightforward generalizations of key equations.

\subsection{{cofe} HEGs}
\label{sec:cof}

\begin{figure}
\includegraphics[width=\linewidth]{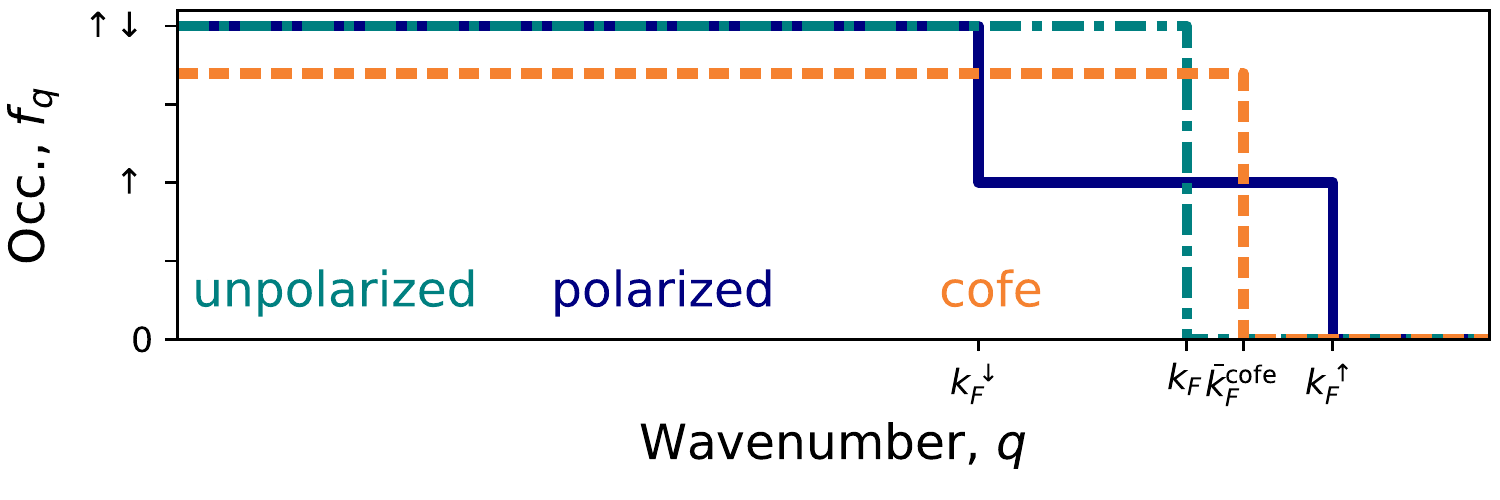} 
\caption{Occupation factors as a function of
wavenumber for an unpolarized gas (dash-dot line),
polarized gas ($\zeta=\half$, solid line)
and cofe gas (dashed line)
-- all at the same electron density.
\label{fig:Occs}}
\end{figure}

With core theory now established, let us proceed
to explore generalizations of HEG physics that
exploit the additional degrees of freedom from
ensembles. Our aim is to develop an understanding of
HEGs that spans ground- and excited-state physics.
To that end, we will reveal the properties of
``constant occupation factor ensemble'' HEGs -- the
meaning of the name will soon become apparent.
The key to generalizations is to invoke both
ground and excited states of HEGs. As we shall
show below, many properties are then uniquely
determined by the occupation factors, $f_q$,
of the HEG; while others depend on $\wv$
explicitly, so require some extra restrictions
on the nature of excited states because
there can be many different sets of weights,
$\wv$, that yield a given $f_q$. 

Eqs~\eqref{eqn:nw}, \eqref{eqn:Tsw} and
\eqref{eqn:Exw} reveal that the density, kinetic
energy and exchange energy of any ensemble
system depend explicitly only on the orbital
occupation factors, $f_i^{\wv}$. In HEGs, we
replace $f_i^{\wv}$ by $f_q$, i.e as a function
of absolute wavenumber, $q$. This follows
from: i) the fact that the KS ``orbitals'' of an HEG are
planewaves $\phi_{\vq}(\vr) \propto e^{i\vq\cdot\vr}$;
and ii) that KS minimization dictates that we fill each
$q=|\vq|$ in full. Thus, given $f_q$ it is possible
to define the density, $n$, as well as the kinetic and
exchange energies.
We will therefore first discuss some HEGs from
the perspective of orbital occupation factors; before
proceeding to refine the definition.

The most intuitive form of HEG is an unpolarized gas in the lowest energy (ground) state.
In orbital (KS) terms, the unpolarized HEG non-interacting ground state is a Slater determinant of doubly occupied plane-wave orbitals.
Occupied states fill in $\up/\down$ pairs up to a single Fermi wave number, $k_F$.
Its wave-number dependent occupation factor and density are,
\begin{align}
f_q^{\unpol}=&2\Theta(k_F-q)\;,
&
k_F=(3\pi^2 n)^{1/3}\;,
\label{eqn:fqunpol}
\end{align}
where $\Theta(x)=\{1\forall x\geq 0;0\forall x<0\}$ is a Heaviside step function.
The density, $n$, of the gas is sufficient to describe the state. 

Ground states realized by exposing the HEG to a uniform external magnetic field (the corresponding vector potential being ignored, as in spin-DFT) have a wave-number dependent occupation factor determined by spin-dependent Fermi wavenumbers,
\begin{align}
f_q^{\pol}=&\Theta(k^{\up}_F-q)+\Theta(k^{\down}_F-q)\;,
~
k_F^{\up,\down}
=(6\pi^2 n_{\up,\down})^{1/3}\;.
\label{eqn:fqpol}
\end{align}
The unpolarized gas is then the special case of $n_{\up}=n_{\down}=\tfrac{n}{2}$ giving $\zeta=0$.
A fully polarized gas has $n_{\up}=n$, $n_{\down}=0$ and $\zeta=1$.
For definiteness, we work under the convention that the majority spin channel is the ``up'' ($\up$) channel.
The density, $n$, and spin-polarization, $\zeta$, are sufficient to describe the state.

In this work, we consider (non-thermal) {\em ensembles} of excited states, which correspond to \emph{averaged} occupation factors. Specifically, we consider ensembles obeying,
\begin{align}
f_q^{\cof}=&\fb\Theta(\kFb^{\cof}-q)\;,
&
\kFb^{\cof}=(6\pi^2 n/\fb)^{1/3}
\label{eqn:fqcof}
\end{align}
The bar on top of $\fb$ (and, thus, $\kFb$) means that this quantity stems from an average w.r.t. an ensemble rather than to a pure state; and `cofe' stands for `constant occupation factor ensemble',~\footnote{`cofe' is pronounced like coffee by the authors.}
reflecting the fact that the system has the same occupation factor right up to a single (ensemble) Fermi level, unlike a polarized gas.
We shall discuss below that the correct interpretation associates $f_q^{\cof}$ with an \emph{unpolarized} ensemble.

Before proceeding further, it is worth considering why we should choose $f_q^{\cof}$ to be constant or zero, rather than any of the infinite number of other options we could have chosen.
The main motivation is simplicity.
Firstly, we aim to keep the number of parameters to two ($n$ and $\fb$) like the spin-polarized gas ($n$ and $\zeta$).
We also aim to ensure that limiting cases (unpolarized and fully polarized gases) are reproduced by cofe gases -- once adapted to inhomogeneous systems the limits respectively correspond to singlet ground states and ground and excited states of one-electron systems.
Finally, noting that both limits have the special feature that they yield constant occupation factors (two and one, respectively), we aim to retain this special feature in between the limits as a sensible generalization that incorporates excited states.
\TG{}{Crucially, Section~\ref{sec:Inhomogeneous} will demonstrate that `cofe' model \emph{can indeed be localized} to approximate inhomogeneous states by recovering meaningful exact conditions.}

\TG{}{The above goals dictate} the form of Eq.~\eqref{eqn:fqcof}, as well as the kinetic and exchange energies of cofe HEGs.
The addition of some extra restrictions (to be discussed below, as needed) on the excited states dictates the remaining properties of cofe-HEGs.
As we shall see later in Section~\ref{sec:Applications} the resulting cofe gas is effective for predicting ground and excited states of inhomogeneous systems.

Figure~\ref{fig:Occs} illustrates the different occupation factors for unpolarized, polarized and cofe HEGs, all at the same density $n$.
The polarized gas has $\zeta=\half$, while the cofe HEG has $\fb=1.7$.
The unpolarized gas has a single Fermi level with double occupations, the polarized gas has two Fermi levels, one higher ($\up$) and one lower ($\down$) than that of the unpolarized gas, and is doubly occupied up to the lower level and then singly occupied to the higher level.
The cofe gas also has a single Fermi level between the unpolarized and $\down$ levels, but is only partly occupied for all $q$.
The choice of $\zeta=\half$ and $\fb=1.7$ ensures that the polarized and cofe HEGs also have the same exchange energy -- as can be seen by evaluating eqs~\eqref{eqn:epsxzeta} and \eqref{eqn:epsxcof}.
\TG{}{We will later exploit this feature in Section~\ref{sec:Correlation}.}

Once we accept to deal with ensembles from constrained occupation factors,  we can mix with equal weights a polarized HEG with its time-reversed partner.
Nothing changes in terms of the evaluation of the energy components.
What changes is the interpretation. 
Now, we can find a continuum of  {\em unpolarized} ensembles of cofe-HEGs, with energies that go from that of the regular unpolarized to that of the regular fully polarized HEGs.
But the ensembles can also accommodate ground states {\em and} excited states (keeping in mind that the polarized gas is itself an excited state in the absence of a magnetic field), in a  sense that will be clarified just below. 

The ingredients of $\Gammah^{\cofe}$ are most easily understood by considering a finite system with four electrons:
\begin{itemize}
\item
The unique unpolarized state is $\iket{\unpol}=\iket{1^2 2^2}$, which is consistent with a Fermi level, $\kFb^{\cof}=\epsilon_2^+$, just above the second orbital energy.
As a singular state we set $w_{\unpol}=1$ and obtain $f_1=f_2=\fb=2$.
\item
The fully polarized system, $\iket{\text{fullpol}}=\iket{1^{\up} 2^{\up} 3^{\up} 4^{\up}}$, is also unique ($w_{\text{fullpol}}=1$).
It has $\kFb^{\cof}=\epsilon_4^+$ (four orbitals allowed) and yields $f_1=f_2=f_3=f_4=\fb=1$.
The corresponding state with all $\down$-electrons has the same energetics (but time-reversed dynamics).
Ensemble averaging the $\up$- and $\down$-spin systems therefore yields a net {\em unpolarized} system with the same energy terms.
\item
But, if we allow three orbitals, we have three maximally polarized ($N^{\up}=3$ and $N^{\down}=1$) states: $\iket{\cof_0}\equiv \iket{1^2 2^{\up} 3^{\up}}$, $\iket{\cof_1}\equiv\iket{1^{\up} 2^2 3^{\up}}$, and $\iket{\cof_2}\equiv \iket{1^{\up} 2^{\up} 3^2}$.
Each state has a spin-polarization $\zeta_{\text{eff}}=\tfrac{3-1}{4}=\half$.
The (non-interacting) Fermi level, $k_F^{\up}$, for $\up$ electrons is always $k_F^{\up}=\epsilon_3^+$. But, we cannot define a level for $\down$ electrons due to holes in $\iket{\cof_1}$ and $\iket{\cof_2}$.
Assigning each of the three states an equal weight, $w_{0}=w_{1}=w_{2}=\tfrac13$, yields $f_1=f_2=f_3=\tfrac43$, as desired.
Thus, $\kFb^{\cof}=k_F^{\up}(=k_F^{\down})=\epsilon_3^+$ (after we also average over spin) for the whole ensemble.
\end{itemize}

\begin{figure}
\includegraphics[width=\linewidth]{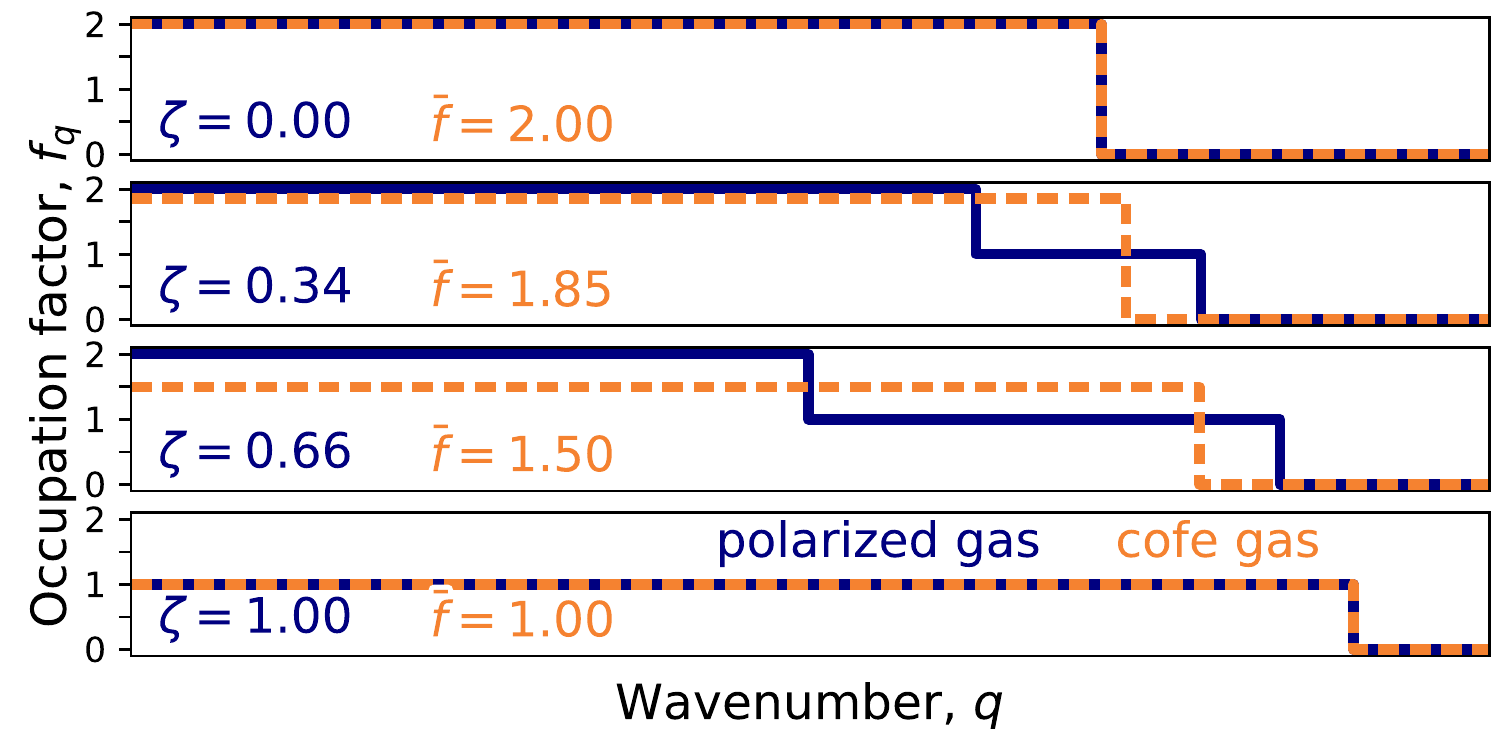} 
\caption{Like Figure~\ref{fig:Occs} except showing
polarized and cofe HEGs at a variety of
$\zeta$ and $\fb$. Note that the polarized
and cofe gas are, as expected, the same for
$\zeta=0$ and $\fb=2$, or $\zeta=1$ and $\fb=1$.
\label{fig:MultiOccs}}
\end{figure}

Replacing orbital indices by $\vq$, and taking the limit $N,V\to\infty$ for fixed density, $n=\tfrac{N}{V}$, and ensemble Fermi level, $\kFb$, yields the actual cof ensemble.
It is composed of ground and excited states all with the same polarization, $\zeta_{\text{eff}}=2/\fb-1$, (and their time reversed partners) where $\fb=6\pi^2 n \kFb^{-3}$ follows from eq.~\eqref{eqn:fqcof}.
Sections~\ref{sec:Hartree} and \ref{sec:Correlation} will expand a little on the specifics of states required for cofe HEGs.
Here and henceforth we drop the superscript from $\kFb^{\cof}$, and simply use $\kFb$.

It is finally worth noting that the energy of a cofe HEG with $\fb=2$ is always equal to that of an \emph{unpolarized} gas with $\zeta=0$, while the energy of a cofe HEG with $\fb=1$ is always equal to that of a \emph{fully polarized} gas with $\zeta=1$ (keeping in mind that the ensemble averages over the time-reversed state).
Figure~\ref{fig:MultiOccs} shows $f_q$ for a selection of polarized and cofe gases between (and at) these limits, all yielding the same density, $n$.
Values of $\zeta$ and $\fb$ are `paired' to yield the same exchange energy -- we will later exploit this pairing in eq.~\eqref{eqn:epscmhd} of Section~\ref{sec:Correlation}.

We will next proceed to compute the energy components of the cofe HEG.
Key results are summarized in Table~\ref{tab:HEGs}.

\subsubsection{Density, kinetic and exchange energies of {cofe HEGs}}

The density,
$n[f_q]:=\int_0^{\infty}f_q \frac{q^2 dq}{2\pi^2}$,
and kinetic energy per particle,
\begin{align}
t_s[f_q]:=&\frac{1}{n[f_q]}\int_0^{\infty}f_q\frac{q^2}{2} \frac{q^2 dq}{2\pi^2}\;.
\end{align}
of an HEG are direct functionals of the
occupation factor distribution, $f_q$.
Prefactors deal with
normalization of the orbitals and energies.
The kinetic energy integral follow from the fact that
$\phi^*_{\vq}(\vr)[-\half \nabla^2\phi_{\vq}(\vr)]
=\half q^2 \phi^*_{\vq}(\vr)\phi_{\vq}(\vr)$.

Typically we are interested in some fixed density,
$n=\tfrac{3}{4\pi r_s^3}$, defined by its Wigner-Seitz
radius, $r_s$, which imposes constraints on $f_q$
(e.g. the Fermi levels in the previous section).
Throughout we will implicitly
define all HEGs to be at fixed Wigner-Seitz radius,
$r_s$, and vary other parameters under this assumption.
Using the occupation factor model for a polarized
gas with fixed $\zeta$ and $r_s$ yields the
kinetic energy given by eq.~\eqref{eqn:tszeta}.

Consider instead a cofe HEG, where
$f_q$ is given by eq.~\eqref{eqn:fqcof}. We obtain,
$n[f_q]=\tfrac{\fb \kFb^3}{6\pi^2}$ from which
we confirm that $\kFb=(6\pi^2 n/\fb)^{1/3}$.
The kinetic energy of a cofe HEG therefore has the
separable expression,
\begin{align}
t_s^{\cof}(r_s,\fb)=&\frac{3\kFb(r_s,\fb)^2}{10}
=t_s(r_s)\bigg[\frac{2}{\fb}\bigg]^{2/3}\;,
\label{eqn:tscof}
\end{align}
using $t_s(r_s)$ from eq.~\eqref{eqn:ts}.

In addition to the density and kinetic energy,
the exchange energy of any HEG may also be
evaluated directly from $f_q$.
Replacing sums over $k$ and $k'$ by integrals
over $\vq$ and $\vq'$ lets us rewrite
Eq.~\eqref{eqn:Exw} as,
\begin{align}
\epsx[f_q] :=& -\frac{1}{n[f_q]}
\int_0^{\infty}\int_0^{\infty}f_{\max(q,q')}V(q,q')
\frac{{q'}^2dq'}{2\pi^2}\frac{q^2dq}{2\pi^2}\;,
\label{eqn:epsxfq}
\end{align}
where, $V(q,q')=\int_{-1}^1 \allowbreak
\frac{\pi dx}{q^2+q'^2-2qq'x}
=\tfrac{\pi}{qq'}\log\tfrac{|q+q'|}{|q-q'|}$
is the spherically averaged Coulomb potential.
A little additional work on the integral
(see Appendix~\ref{app:Functionals} for details)
yields eq.~\eqref{eqn:epsxzeta} for a polarized gas;
and,
\begin{align}
\epsx^{\cof}(r_s,\fb)=&-\frac{\fb}{n}
\int_0^{\kFb}\frac{q}{\pi}\frac{q^2dq}{2\pi^2}
=\epsx(r_s)\bigg[\frac{2}{\fb}\bigg]^{1/3}\
\;,
\label{eqn:epsxcof}
\end{align}
for cofe HEGs, where $\epsx(r_s)$ is the
unpolarized HEG expression of eq.~\eqref{eqn:epsx}.

Although not necessary for computing, $\epsx$,
we may similarly derive an expression for the
HEG exchange hole, defined in eq.~\eqref{eqn:n2x}.
We obtain,
\begin{align}
n_{2,\xrm}^{\cof}(R;r_s,\fb)=&\Pi_{\xrm}^{\cof}(r_s,\fb)
N(\kFb R)
\label{eqn:n2xcof}
\end{align}
where,
\begin{align}
\Pi_{\xrm}^{\cof}(r_s,\fb)=&
-\fb\int_0^{\kFb}\frac{q^3}{3\pi^2}\frac{q^2dq}{2\pi^2}
=\frac{-n^2}{\fb}
\label{eqn:Pixcof}
\end{align}
is the on-top pair-density of the exchange
hole, and $N(x):=9[\sin(x)-x\cos(x)]^2/x^6$
is a function. We will use the relationship between
the exchange energy and exchange hole
to help in deriving the properties of the
Hartree energy, in the next section.

\subsubsection{Hartree energy of {cofe HEGs}}
\label{sec:Hartree}

The ensemble Hartree energy functional is given in
eq.~\eqref{eqn:EHw}.
This term is usually ignored in HEG discussions because
$n_{2,\Hrm}=n^2$ in polarized and unpolarized gases
at arbitrary $\zeta$, which means that $\epsilon_{\Hrm}$
exactly cancels the energy of the positive background charge,
$\epsilon_{\bg}$ -- that is,
$\epsilon_{\Hrm}[n_{2,\Hrm}=n^2]=-\epsilon_{\bg}$.
In cofe HEGs this cancellation is incomplete. The
singular background charge is guaranteed, by
charge neutrality, to be cancelled in full. 
However, the Hartree pair-density, $n_{2,\Hrm}^{\wv}\neq n^2$,
differs from the background charge density, $n^2$,
and thus $\epsilon_{\Hrm}^{\wv}$ includes
additional terms. The
energy per particle of an ensemble HEG is,
\begin{align}
e^{\wv}[f^{\wv}_q]
:=t_s[f^{\wv}_q] + \depsH^{\wv}[f^{\wv}_q]
+ \epsx[f^{\wv}_q] + \epsc^{\wv}[f^{\wv}_q]\;,
\label{eqn:epsw}
\end{align}
where superscripts $^{\wv}$ indicate an explict
dependence on the nature of the ensemble.
The additional positive Hartree energy contribution,
\begin{align}
\depsH^{\wv}=\epsilon_{\Hrm}^{\wv}-
\epsilon_{\bg}=\frac{1}{n}\int \Delta n_{2,\Hrm}^{\wv}(R)
\frac{d\vR}{2R}\;,
\end{align}
may be evaluated [eqs~\eqref{eqn:Ewn2} and \eqref{eqn:n2H}]
using the ensemble Hartree pair-density deviation,
$\Delta n_{2,\Hrm}^{\wv}=n_{2,\Hrm}^{\wv}-n^2$.

We therefore seek closed-form expressions
for $n_{2,\Hrm}^{\cof}$ and $\depsH^{\cof}$
for the special case of a cofe HEGs with
maximal polarization within the ensemble,
as defined earlier.
Full details for Hartree expressions are rather
involved so have been left to Appendix~\ref{app:Hartree}.
The rough argument is as follows:
1) the background charge is cancelled by $\FE=\FE'$
terms in \eqref{eqn:EHw} or \eqref{eqn:n2H},
so we need only evaluate $\FE\neq \FE'$ terms;
2) the cof ensemble states, $\iket{\FE}$,
contain every possible combination of paired and
unpaired orbitals up to $\kFb$;
3) each of these states is weighted equally;
4) we may therefore use combinatorial arguments
to evaluate key expressions.
The final step recognises that each state may
be defined by a set, $\{\vq\}_{\text{double}}$,
of doubly occupied orbitals, such that the remaining
occupied orbitals (with $|\vq|\leq \kFb$)
contain only an $\up$ electron. Each
non-interacting state is then a Slater
determinant consistent with the occupations,
whose properties may be understood via
$\{\vq\}_{\text{double}}$ and $\kFb$.

Appendix~\ref{app:Hartree} yields, $\depsH^{\cof}(r_s,\fb):=
\tfrac{C_{\Hrm}}{r_s}\tfrac{(2-\fb)(\fb-1)}{\fb^{4/3}}$,
[eq.~\eqref{eqn:depsHinApp}]
where $C_{\Hrm}=2^{1/3}C_{\xrm}$. We rewrite this as,
\begin{align}
\depsH^{\cof}(r_s,\fb)
=|\epsx(r_s,\fb)|\frac{(2-\fb)(\fb-1)}{\fb}\;,
\label{eqn:depsHcof}
\end{align}
for use in eq.~\eqref{eqn:epsw} and later expressions.
This result follows from the fact that,
\begin{align}
n_{2,\Hrm}^{\cof}(R;r_s,\fb)
=&n^2+\Delta\Pi_{\Hrm}^{\cof}(r_s,\fb)N(\kFb R)\;,
\\
\Delta\Pi_{\Hrm}^{\cof}(r_s,\fb)=&n^2\tfrac{(2-\fb)(\fb-1)}{\fb^2}
=-\tfrac{(2-\fb)(\fb-1)}{\fb}\Pi_{\xrm}^{\cof}\;,
\end{align}
where $N(x)$ is the same expression used in
\eqref{eqn:n2xcof}.

\subsubsection{Energies in the low-density limit of {cofe HEGs}}
\label{sec:ldCorr}

We cannot analytically evaluate the energy
of an HEG at arbitrary density, $n$. We can,
however, semi-analytically evaluate it
in the high density (large $n$, small $r_s$)
and low density (small $n$, large $r_s$) limits.
The high density limit may be obtained from
a series solution around the Kohn-Sham solution.
In the low density limit, the electrons are far
enough apart to undergo a process known as
a Wigner crystallisation.~\cite{Wigner1934,Wigner1938}
The resulting ``strictly correlated electron'' physics
may then be understood via a classical leading
order term, with quantum corrections.
The transition occurs at $r_s\approx 100$~Bohr.

Recent work~\cite{Gould2023-ESCE} has shown that any dependence on ensemble properties must vanish in the low-density limit of any finite system;
so that \emph{all excited state properties become degenerate to both leading and sub-leading order}.
It is very likely that this result also holds true in the thermodynamic limit of HEGs, as justified by the following intuition:
\begin{enumerate}
\item As the density becomes small, the distance between electrons becomes large and the particles become effectively classical with a quantum state defined by fluctuations around a classical minima;
\item Whether the system is finite, or infinite, the fluctuations may be ``excited'' any number of times with no impact on the \emph{classical} leading order term of the interaction energy;
\item Furthermore, the next leading order \emph{quantum} correction from zero-point energy fluctuations around the classical minima are dictated only by the density constraint, and are therefore also independent of excitation structure. 
\end{enumerate}

This result has important implications for both spin-polarized and cofe HEGs, as both may be represented as ensemble of excited states -- with specific properties governed by $\zeta$ or $\fb$, respectively.
It follows from the above that the leading two orders of their low-density energies are independent of the excitation structure.
Consequently, energies are independent of $\fb$ and $\zeta$.
Independence of $\zeta$ has long been theorized for spin-polarized HEGs.
Recent QMC data~\cite{Azadi2022} provides confirmation of this result.

\newcommand{\ld}{\text{ld}}
\newcommand{\hd}{\text{hd}}
\newcommand{\mhd}{\text{mhd}}

The leading order terms correspond to $1/r_s$ and $1/r_s^{3/2}$ in the usual large-$r_s$ series description of HEGs.
Therefore, ensemble and spin effects can only contribute at $O(1/r_s^2)$.
The (Hartree) exchange- and correlation energy of strictly correlated electrons in the low-density limit (ld) therefore obeys $\lim_{r_s\to\infty}\epsilon_{\Hxc}(r_s,\zeta)=\epsilon_{\Hxc}^{\ld}(r_s)$, where,
\begin{align}
\epsilon_{\Hxc}^{\ld}(r_s):=\frac{-C_{\infty}}{r_s}
+ \frac{C'_{\infty}}{r_s^{3/2}} + \ldots\;,
\label{eqn:epsHxcld}
\end{align}
includes only the part of the Hartree energy that is not cancelled by background charge.
The best estimates for coefficients are $C_{\infty}=0.8959\approx 1.95 C_{\xrm}$ and $C'_{\infty}=1.328$.~\cite{Alves2021,Smiga2022}

In regular HEGs, the Hartree term is fully cancelled by background charge so can be ignored.
For polarized HEGs we therefore obtain, $\lim_{r_s\to\infty}\epsc=\epsilon_{\Hxc}^{\ld}-\epsx$, and,
\begin{align}
\lim_{r_s\to\infty}
\epsc(r_s,\zeta):=\frac{-C_{\infty}+C_{\xrm}f_{\xrm}(\zeta)}{r_s}
+ \frac{C'_{\infty}}{r_s^{3/2}} + \ldots\;,
\end{align}
using $f_{\xrm}$ from eq.~\eqref{eqn:epsxzeta}.
By contrast, in a cofe HEG there is a non-zero component ($\depsH^{\cof}$) in the Hartree energy.
It therefore follows that, $\lim_{r_s\to\infty}\epsc^{\cof}=\epsxc^{\ld} - \depsH^{\cof} - \epsx^{\cof}$.
We finally obtain,
\begin{align}
\lim_{r_s\to\infty}
\epsc^{\cof}(r_s,\fb)=&\frac{-C_{\infty}
+ C_{\xrm}f_{\Hx}^{\cof}(\fb)}{r_s}
+ \frac{C'_{\infty}}{r_s^{3/2}} + \ldots\;,
\end{align}
where,
\begin{align}
f_{\Hx}^{\cof}(\fb)=\bigg[\frac{2}{\fb}\bigg]^{1/3}
\frac{(\fb-1)^2+1}{\fb}\;,
\label{eqn:fHxcof}
\end{align}
follows from eqs~\eqref{eqn:epsxcof} and \eqref{eqn:depsHcof}.
This is the appropriate low-density series expansion for
the correlation energy of cofe HEGs.

\subsubsection{State-driven correlation energies of {cofe HEGs}}
\label{sec:Correlation}

In general, the correlation energy of an ensemble is
separable into two terms,~\cite{Gould2019-DD,Fromager2020-DD,Gould2020-FDT}
\TG{}{$E_{\crm}^{\wv}:=E_{\crm}^{\SD,\wv}+E_{\crm}^{\DD,\wv}$ [eq.~\eqref{eqn:Ecw}]} where each covers different physics of the ensemble.
The ``state-driven'' (SD) term is the \emph{only term present} in pure states, such as polarized gases.
In general ensembles, it is like a weighted average of conventional correlation energies for the different states of
the ensemble.
The ``density-driven'' (DD) term reflects the fact that the densities of the individual Kohn-Sham and interacting states that form the ensemble are not necessarily the same -- only the averaged ensemble density is the same.

We expect that only the SD part of the correlation energy should form part of the xc energy used in density functional approximations, so focus here on this term -- we explain this choice in Section~\ref{sec:Inhomogeneous}.
Our goal is therefore to determine $\epsc^{\SD,\cof}(r_s,\fb)$ as a function of $r_s$ and $\fb$, which we will use as a basis for parameterization in the next section.
This involves considering the high- and low-density limits of matter (and therefore cofe-HEGs), for which exact results will be derived.
We will also discuss how to repurpose existing data for values in between these limits.
Comprehensive analysis of both state- and density-driven correlation terms is reported in Appendix~\ref{app:CorrAll}.
Below, we summarize key elements of the SD correlation energy analysis.

The division into SD and DD terms is not unique,%
~\cite{Gould2019-DD,Fromager2020-DD,Gould2020-FDT}
and any explicit study of the separation into SD and DD terms requires accessing the properties of a variety of excited states of \emph{interacting} HEGs.
Nevertheless, discussion near eq.~(14) of \rcite{Gould2020-FDT} argues that the SD correlation energy may be written in adiabatic connection and fluctuation-dissipation theorem (ACFD) form:
\begin{align}
\epsc^{\SD}:=&\frac{1}{n}\int_0^1d\lambda
\int_0^{\infty}\frac{-d\omega}{\pi}
\int \frac{d\vr d\vrp}{2|\vr-\vrp|}
\nonumber\\&\times
\big[\chi_{\lambda}(\vr,\vrp;i\omega)-\chi_0(\vr,\vrp;i\omega)\big]\;,
\label{eqn:epscFDT}
\end{align}
\TG{}{which is Eq.~\eqref{eqn:EcSDw} adapted to HEGs.}
Here, $\chi_0$ is the collective density-density response of the non-interacting cofe HEG defined earlier -- i.e. the ensemble of part-polarized ground- and excited states that yield $\fb$.
$\chi_{\lambda}$ is its equivalent for a scaled Coulomb interaction $\tfrac{1}{R}\to \tfrac{\lambda}{R}$.
In principal, the individual states in the interacting ensemble may be followed from their known $\lambda=0$ values to their unknown value at arbitrary $\lambda$, although this is not required in practice.

The key step toward understanding how to separate
and parametrise terms is to use the
random-phase approximation (RPA).
RPA becomes exact (to leading order) in the
high-density limit.~\cite{PW92} More generally,
RPA provides an approximate solution for
eq.~\eqref{eqn:epscFDT}, and thus provides insights
into the SD correlation term. Details are provided
in Appendix~\ref{app:CorrRPA}.
Key findings are: i) that, $\epsc^{\SD,\cof}$
is approximately linear in $\fb$ for high densities;
ii) for low densities we obtain a scaling that is
similar to $f_{\xrm}^{\cof}(\fb)$. 
Appendix~\ref{app:CorrGen} then
uses the RPA results, and fundamental theory,
to argue that,
\begin{align}
\epsc^{\SD,\cof,\hd}(r_s,\fb)=&
(\fb-1)\epsc(r_s,0)
+ (2-\fb)\epsc(r_s,1)\;,
\label{eqn:epschd}
\\
\epsc^{\SD,\cof,\ld}(r_s,\fb) =&
\epsx(r_s)\big[\tfrac{C_{\infty}}{C_{\xrm}} - f_{\xrm}^{\cof}(\fb)\big]
+ \tfrac{C_{\infty}'}{r_s^{3/2}}\;,
\label{eqn:epscld}
\end{align}
are, respectively, the exact high- and
low-density limits of $\epsc^{\cof}$.
That is, $\lim_{r_s\to 0}\epsc^{\SD,\cof}(r_s,\fb)
=\epsc^{\SD,\cof,\hd}(r_s,\fb)$
and $\lim_{r_s\to \infty}\epsc^{\SD,\cof}(r_s,\fb)
=\epsc^{\SD,\cof,\ld}(r_s,\fb)$.

Filling in the gaps between these limits requires quantum Monte Carlo (QMC) calculations:
which, however, are \emph{only available for spin-polarized ground-states} of homogenous gases.
Appendix~\ref{app:CorrGen} therefore shows how to reuse existing spin-polarized QMC data for the in-between regime, by adapting it for cofe HEGs.
Specifically, it argues that,
\begin{align}
\epsc^{\SD,\cof,\mhd}(r_s,\fb)\equiv&
\epsc^{\QMC}(r_s, \zeta=\fmap^{-1}(\fb))
\label{eqn:epscmhd}
\end{align}
is a reasonable {\em approximation} for medium-high densities (mhd).
The key assumption behind this relationship is that HEGs with same exchange energy should have a similar state-driven correlation energy.
Thus, $\fmap(\zeta)$ is a function yielding, $\epsx^{\cof}(r_s,\fmap(\zeta))=\epsx(r_s,\zeta)$ and $\epsx^{\cof}(r_s,\fb)=\epsx(r_s,\fmap^{-1}(\fb))$.
Eq.~\eqref{eqn:epscmhd} becomes exact in the low-density limit, but incorrect in the high-density limit. 
More information is provided in Appendix~\ref{app:CorrGen}. 

\begin{figure}
\includegraphics[width=\linewidth]{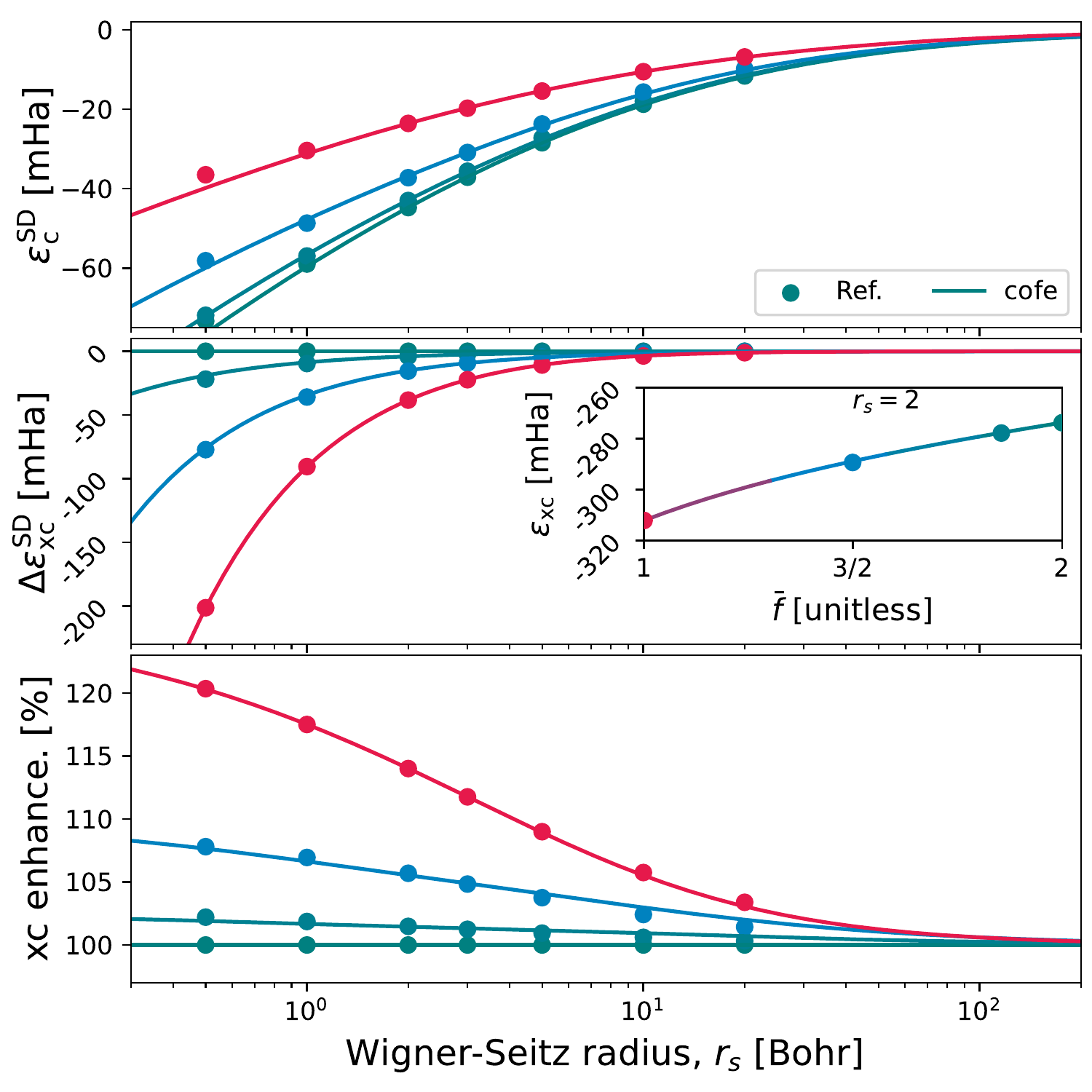} 
\caption{Correlation (top) and xc (middle) energies and xc (bottom) enhancement factors for HEGs as a function of $r_s$ and $\fb\in (2,1.85,1.50,1)$.
Plots show the cofe (solid lines) parametrisation introduced here, and the adapted benchmark results from \rcite{Spink2013} (circles).
The inset plot shows $\epsxc$ (cofe and benchmark) as a function of $\fb$, for $r_s=2$.
Line colours indicate the value of $\fb$ (see inset for values).
\label{fig:epsxcParam}}
\end{figure}

Appendix \ref{app:Param} details parametrization of $\epsc^{\SD,\cof}(r_s,\fb)$ for arbitrary densities, based on the theoretical work in this section.
As an intermediate step, it also introduces approximations for $\fmap$ and its inverse, for use in eq.~\eqref{eqn:epscmhd}.
Key results are visually summarized in Figure~\ref{fig:epsxcParam}, which compares the parametrisation of $\epsc^{\SD,\cof}$ with the (adapted) reference data used to fit it.
The top plot shows correlation energies, $\epsc^{\SD,\cof}(r_s,\fb)$.
The middle plot shows deviations, $\Delta\epsxc^{\SD,\cof}=\epsxc^{\SD,\cof}(r_s,\fb)-\epsxc^{\SD,\cof}(r_s,2)$, from unpolarized gas values.
The bottom plot shows xc enhancement factors, $\epsxc^{\SD,\cof}(r_s,\fb)/\epsxc^{\SD,\cof}(r_s,2)$, which must approach one (100\%) in the low-density (large $r_s$) limit.

\section{From {cofe HEGs} to real systems}
\label{sec:Inhomogeneous}

\TG{}{As discussed in Sec.~\ref{sec:LDA}, Kohn and Sham used an inhomogeneous description of the (quantum mechanical) kinetic and (classical) Hartree energies, together with an HEG-based approximation for the exchange and correlation energy only, per Eq~\eqref{eqn:LDA}.
It is natural to assume that ensembles and excited states can benefit from a similar treatment.
However, for ensembles it is important to work with the corresponding {\em extended} functionals and, in particular, the Hartree functional forms described in Eq.~\eqref{eqn:EHw}.
Recent applications have confirmed the advantages of using \eqref{eqn:EHw}, rather than the traditional `classical' electrostratic energy.~\cite{Gould2020-Molecules,Gould2021-DoubleX,Gould2022-pEDFT}}

\TG{}{Switching to ensemble of excited states, and mimicking Eq~\eqref{eqn:LDA} to employ the cofe HEGs, leads to,}
\begin{align}
E_{\eLDA}^{\wv}=&\min_n\bigg\{
T_s^{\wv}[n] + \int n(\vr) v(\vr)d\vr + E_{\Hrm}^{\wv}[n]
\nonumber\\&
+ \int n(\vr)\epsilon_{\xc}^{\cofe}\big(r_s(\vr),\fb(\vr)\big)d\vr
\bigg\}\;.
\label{eqn:eLDA}
\end{align}
Here, we have replaced pure state $T_s$ and $E_{\Hrm}$ by their ensemble equivalents, $T_s^{\wv}$ [eq.~\eqref{eqn:Tsw}] and $E_{\Hrm}^{\wv}$ [eq.~\eqref{eqn:EHw}]; and locally approximated the xc energy by the cofe LDA with \emph{local} Wigner-Seitz radius, $r_s(\vr)$, and \emph{local} effective occupation factor, $\fb(\vr)$.
\TG{}{Note, as was done earlier we leave $^{\wv}$ superscripts off local quantities ($n$, $r_s$ and $\fb$).}

\newcommand{\SE}{|\FE\rangle}
\newcommand{\Ph}{\hat{P}}

\TG{}{Furthermore -- and much more usefully in practice -- it is possible to use Eq.~\eqref{eqn:eLDA} to generate state-resolved energies, $E_{\SE}$, for target states, $\SE$,%
~\footnote{Or level specific contributions, in case of degeneracies. To keep notation and illustrations simple, as done also before, we shall not explicate all the details.}
that obey stationary conditions.~\cite{Gould2024-Stationary}
The resulting eLDA energy expression is:
\begin{align}
E^{\eLDA}_{\SE}=&T_{s,\SE}[n_{\SE}] + \int n_{\SE}(\vr) v(\vr) d\vr + E_{\Hrm,\SE}[n_{\SE}]
\nonumber\\&
+ \int n_{\SE}(\vr)\epsilon_{\xc}^{\cofe}\big(r_{s,\SE}(\vr),\fb_{\SE}(\vr)\big)d\vr
\;.
\label{eqn:eLDA_State}
\end{align}
In brief, Eq.~\eqref{eqn:eLDA_State} follows from applying the exact weight-derivative relationship, $E_{\SE}=\Dp{w_{\FE}}E^{\wv}$, [for any positive weight, $w_{\FE}$, in the ensemble --- see \eqref{eqn:Ew}] to the eLDA energy expression, after assuming that orbitals are fixed in the derivative.~\cite{Gould2021-DoubleX}
$T_{s,\SE}$ and $E_{\Hrm,\SE}$ are obtained from the inhomogeneous system. 
We approximate the exchange and state-driven correlation energy via,
\begin{align}
\epsilon_{\xc}^{\cofe}(r_s,\fb):=\epsx^{\cof}(r_s,\fb)
+ \epsc^{\SD,\cof}(r_s,\fb)\;,
\label{eqn:epsxccofe}
\end{align}
from the local Wigner-Seitz radius, $r_{s,\SE}(\vr)$, and local effective occupation factor, $\fb_{\SE}(\vr)$.
We will discuss energy terms in the next subsection and return to $\fb$ in Section~\ref{sec:fbar}.
}

\subsection{State-resolved treatment}
\label{sec:Exact}

\TG{}{
In the state-resolved treatment, each excited state, $\SE$, is dictated by a set of orbital occupancies and spin-symmetry (singlet, doublet, triplet, etc) -- in weakly-correlated wave functions these would indicate the dominant Slater Determinants in a configuration expansion of the true excited state.
Both $T_{s,\SE}$ and $n_{s,\SE}$ take their usual forms, but with orbital occupation factors, $\theta_i^{\SE}$, taken from the excited state.
The Hartree energy is obtained from Eq.~\eqref{eqn:EHw} to yield,
\begin{align}
E_{\Hrm,\SE}:=\Dp{w_{\FE}}E_{\Hrm}^{\wv}\equiv U[n_{\SE}] + 2\sum_{\FE'<\FE}U[n_{s,\FE\FE'}]\;.
\label{eqn:EHstate}
\end{align}
The first term is the typical ground state Hartree energy, $E_{\Hrm}[n_{\SE}]=U[n_{\SE}]$, but the extra $U[n_{s,\FE\FE'}]$ terms involve the Coulomb energy of a transition density from KS state, $\iket{\FE_s}$ to a lower energy KS state, $\iket{\FE'_s}$, of the same symmetry, i.e ``de-excitations''.
For the lowest-lying excitation of each given symmetry (spatial or spin) a standard-looking pure state problem is well-defined, but 
complications appear when considering higher excitations.
}

\TG{}{
This work considers only the lowest energy excitation for each allowed symmetry, and some other symmetry-protected excited states, and thus solves Eq.~\eqref{eqn:eLDA_State} via minimization.
Specifically, it considers three types of excited states, whose KS states may be fully defined via electron promotion ($\Ph_{\text{from}}^{\text{to}}$) from the lowest energy doubly occupied singlet state, $\iket{S_0}$:
single excitation (promotion) to a triplet,
$$\iket{T_i^a}\equiv \tfrac{1}{\sqrt{2}}[\Ph_{i\up}^{a\up}-\Ph_{i\down}^{a\down}]\iket{S_0}\;;$$
single excitation (promotion) to a singlet,
$$\iket{S_i^a}\equiv \tfrac{1}{\sqrt{2}}[\Ph_{i\up}^{a\up}+\Ph_{i\down}^{a\down}]\iket{S_0}\;;$$
double excitation (promotion) to a singlet,
$$\iket{S_{i^2}^{a^2}}\equiv \Ph_{i\up}^{a\up}\Ph_{i\down}^{a\down}\iket{S_0}\;.$$
Thus, once we define each state via $i$, $a$, and the nature of the excitation, we are able to find the eLDA energy by minimizing Eq.~\eqref{eqn:eLDA_State} with respect to orbitals using the approach detailed in \SMSec{II}.
Orbital self-consistent solutions, $\{\phi_i^{\SE}\}$, are therefore different in each state, $\SE$.
}


\TG{}{
The xc energy approximation is detailed in Appendix~\ref{app:Param}.
The exchange energy term takes the exact (for cofe gases) form,
\begin{align}
\epsx^{\cof}(r_s,\fb):=&
\epsx(r_s)[2/\fb]^{1/3}
\end{align}
while the state-driven correlation energy term may be parametrized as,
\begin{align}
&\epsc^{\SD,\cof}(r_s,\fb):=(\fb-1)\epsc^0+(2-\fb)\epsc^1
\nonumber\\&
+(\fb-1)(2-\fb)\big[
M_2(r_s)
+(\tfrac32-\fb)M_3(r_s) \big]
\end{align}
for $\epsc^{\zeta}$ computed using eq.~\eqref{eqn:FPW92} (parameters in Table~\ref{tab:Params}).
Here, $M_{2,3}(r_s)$ involve weighted sums (coefficients in Table~\ref{tab:Weights}) over functions, $\epsc^{\zeta}$.
}

\TG{}{
Finally, we stress that Eq.~\eqref{eqn:eLDA_State} ignores density-driven (DD) correlations entirely, i.e. sets $E_{\crm}^{\DD,\wv}\equiv 0$.
This is because DD correlations, like Hartree interactions (either regular or ensemblized), are highly non-local quantities [see \eqref{eqn:n2cDD}] that must not be treated via {\em any} LDA.
Ignoring DD-correlation should not significantly affect our conclusions, as supported by the close agreement between TDLDA and eLDA on single particle excitations in Sec.~\ref{sec:Applications}.}

\subsection{Effective occupation factor}
\label{sec:fbar}

\TG{}{
The final step toward the eLDA is to justify and derive a {\em local} effective occupation factor.
Eq.~\eqref{eqn:eLDA_State} assumes that exchange and state-driven correlations may be approximated locally.
In fact, we know from recent works~\cite{Gould2021-DoubleX,Gould2022-pEDFT} that excited state physics \emph{can be modelled effectively} by re-using ground-state semi-local xc approximations, because applying combination rules~\cite{Gould2021-DoubleX} to existing DFAs can be very effective in practice.
Physically, this may be justified by recognising that exchange and state-driven correlations are response-based properties~\cite{Gould2020-FDT} and are thus consistent~\footnote{This relationship follows from the fluctuation-dissipation theory for DFT, as it relates xc pair-densities (and thus xc holes) to frequency-integrals over density-density response functions.~\cite{Gould2020-FDT} } with `xc hole'-based approximations.~\cite{PribramJones2015}
}

\begin{figure}
   \includegraphics[width=\linewidth]{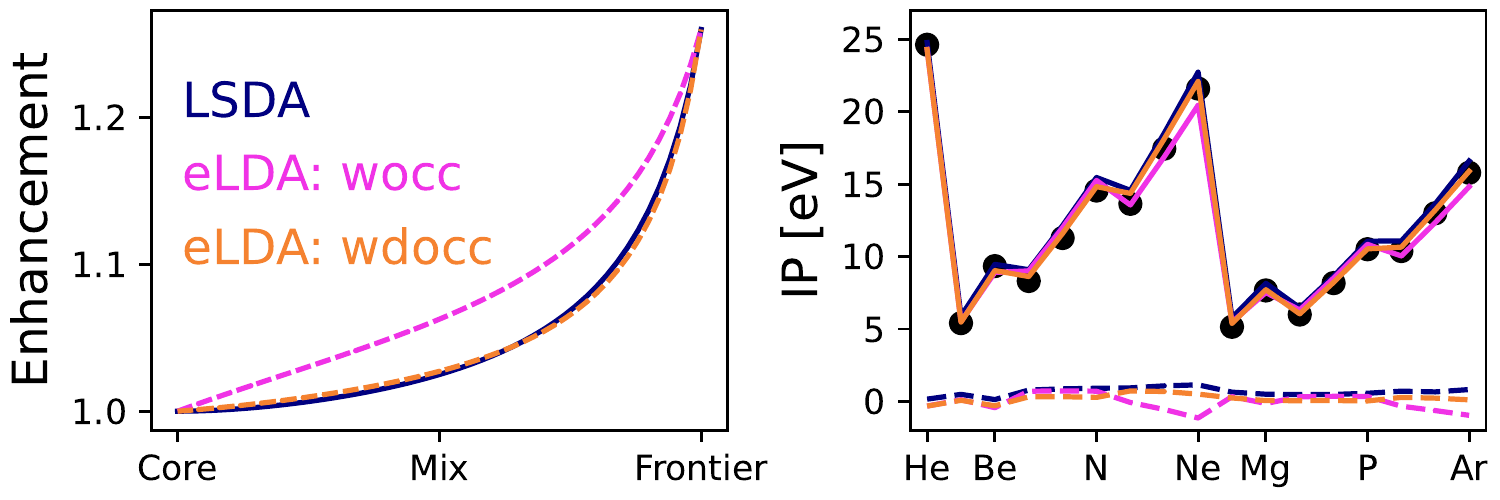} 
   \caption{
   {\bf Left}: Doublet enhancement factor for different ratios of core and frontier occupied
   orbitals from LSDA (teal);
   and eLDA with effective $\fb$ from Eq.~\eqref{eqn:fbar_wocc} (wocc, magenta)
   and Eq.~\eqref{eqn:fbar_inh} (wdocc, orange).
   {\bf Right}: Ionisation potentials  (IPs) for atoms He--Ar using a conventional
   LSDA~\cite{VWN} (navy), Eq.~\eqref{eqn:fbar_wocc}  (magenta)
   and Eq.~\eqref{eqn:fbar_inh} (orange).
   Dashed lines indicate deviations from experimental IPs.
   \label{fig:fbar}
   }   
\end{figure}

\TG{}{
We therefore see that even the regular LDA (and  extensions) \emph{can} be reused.
Thus it is natural to expect that the cofe gas -- which accounts for the excited states of jellium -- can do even better.
{\em But how do we localize the energy of the cofe gas?}
The first step toward an answer is to recognise that the exact exchange energy density of ensembles or excited states depends only on the orbitals, $\{\phi_i\}$, and their occupations, $\{f_i\}$,~\cite{Gould2020-FDT}
which can be seen by rewriting Eq.~\eqref{eqn:Exw} as $E_{\xrm}^{\wv}:=\int n(\vr)\epsx^{\text{exact}}(\vr,\{f_i^{\wv}\})d\vr$ where,
\begin{align}
&\epsx^{\text{exact}}(\vr,\{f_i\}):=
-\frac12 \sum_i \frac{f_i n_i(\vr)}{n(\vr)}
\nonumber\\&\times
\bigg\{ v_U[n_i](\vr)
+2\Re\sum_{i'<i} \frac{n_{ii'}(\vr)}{n_i(\vr)}
v_U[n_{i'i}](\vr)
\bigg\}\;,
\label{eqn:epsxExact}
\end{align}
using $n_{ii'}=\phi_i\phi_{i'}^*$ and $v_U[n]=\int n(\vrp)\tfrac{d\vrp}{|\vr-\vrp|}$.
Furthermore, because combination rule EDFAs are effective for x and c, and depend only on local densities,~\cite{Gould2021-DoubleX} we know we can use orbital densities, $n_i=|\phi_i|^2$, instead of orbitals.
Thus, $\epsxc(\vr)\approx\epsxc(\{n_i(\vr)\},\{f_i\})$, is a viable approximation.
}

\TG{}{
The second step is to consider limiting cases.
The motivation behind cofe gases (per Section~\ref{sec:cof}) is to replace $\zeta$ by $\fb$ as an interpolative variable between unpolarized and fully polarized physics, as the simplest extension of HEGs with minimal bias to spin-physics.
By construction, this model should naturally capture two limiting cases for eLDAs: i) LSDA should be reproduced for the doublet ground state xc energy of any one-electron system (where $\zeta=1$ and $\fb=1$ are unambiguous); ii) LDA should be reproduced in unpolarized ground states (where $\zeta=0$ and $\fb=2$ are unambiguous).
Another natural limiting case is the cofe HEG itself.
We can capture these limiting cases by writing $\epsxc(\{n_i(\vr)\},\{f_i\})\approx \epsxc^{\cofe}(n(\vr),\fb(\vr))$ where $\fb(\vr)=\fb(\{f_i\},\{n_i(\vr)\})$ is a local mapping (approximation) for ground and excited states that is constrained in form to reproduce known limits.}

\TG{}{Thus, we make an ansatz for $\fb(\{f_i\},\{n_i(\vr)\})$ that is correct, by construction, for limiting cases but that -- as we will show below -- is also effective for inhomogeneous systems.
As anticipated at the beginning of this Section, it is convenient to switch from a state-average to a state-resolved approach.
A first natural ansatz is thus the density-weighted average of occupation factors,
\begin{align}
\fb_{\text{wocc}}^{\SE}(\vr):=&
\sum_i \theta_i^{\SE} \frac{\theta_i^{\SE} n_i(\vr)}{n(\vr)}
=\frac{\sum_i (\theta_i^{\SE})^2 n_i(\vr)}{\sum_i \theta_i^{\SE} n_i(\vr)}\;,
\label{eqn:fbar_wocc}
\end{align}
expressed here for a given excited state, $\SE$ (we replace $\theta_i^{\SE}$ by $f_i^{\wv}$ in ensembles).
Here, $n_i=|\phi_i|^2$ is the density of orbital $\phi_i$, and $\theta_i^{\SE}$ is its occupation factor in $\SE$.}
It is easily verified that this ansatz is exact for \TG{}{the cofe HEG, unpolarized and one-electron cases}, so is \emph{prima facie} a reasonable extension to inhomogeneous systems.
However, testing (to be discussed below) reveals that this ansatz can yield poor results for ground states.
These errors come from the effective spin-enhancement being too great in regions that are partly-polarized [i.e. where $1<\fb(\vr)<2$].
Fortunately, we may exploit the fact that there are other choices of inhomogeneous $\fb(\vr)$ that yield correct \TG{}{limits}, but that do not hamper performance in inhomogeneous ground states.

We therefore (see \SMSec{I} for details) instead adopt \TG{}{an empirical} double weighted average,
\begin{align}
\fb_{\text{dwocc}}^{\SE}(\vr):=
\frac{\sum_i (\theta_i^{\SE})^{1/3} n_i(\vr)}{\sum_i \theta_i^{\SE}n_i(\vr)}
\frac{\sum_i (\theta_i^{\SE})^{8/3} n_i(\vr)}{\sum_i \theta_i^{\SE}n_i(\vr)}\;,
\label{eqn:fbar_inh}
\end{align}
for calculations.
\TG{}{Here, in addition to satisfying exact limits, eq.~\eqref{eqn:fbar_inh} also approximately replicates the LSDA in general doublet systems.
Compliance with this constraint is justified by the fact that eq.~\eqref{eqn:epsxExact} is exact for spin-polarized ground states,~\cite{Gould2021-DoubleX} which therefore provides a norm for general excited state physics.
}

The left panel of Figure~\ref{fig:fbar} illustrates the importance of choosing $\fb$ appropriately.
It shows the exchange enhancement factor of a doublet system (density $n=2n_{\text{Core}}+n_{\text{Frontier}}$ \TG{}{so that $\theta_{\text{Core}}=2$ and $\theta_{\text{Frontier}}=1$}), for different ratios of $n_{\text{Frontier}}/n_{\text{Core}}$, using standard spin-polarization, and ensemble enhancement with Eqs~\eqref{eqn:fbar_wocc} and \eqref{eqn:fbar_inh}.
It is clear that \eqref{eqn:fbar_wocc} over-enhances exchange in general, relative to LSDA.
By contrast, \eqref{eqn:fbar_inh} matches quite closely to the spin-polarized enhancement of LSDA for all ratios.

\section{Applications}
\label{sec:Applications}

\subsection{Ground states}

How well does eLDA work in practice?
The next section will address excited state energies.
{\em But, first, we need to ensure that the eLDA does not make things worse for ground state energies}.
The right panel of Figure~\ref{fig:fbar} therefore shows the ionization potentials (IPs) of atoms -- that is
the difference in ground state energies between the atom and its cation -- computed with Eq.~\eqref{eqn:eLDA} using Eq.~\eqref{eqn:fbar_wocc} and Eq.~\eqref{eqn:fbar_inh}.
IPs provide a useful test of $\fb(\vr)$ on ground states because the occupation factors of atoms and ions are always different and at least one system always involves an unpaired electron.

The figure reveals that Eq.~\eqref{eqn:fbar_inh} yields results that are consistently close to standard LSDA
calculations, whereas \eqref{eqn:fbar_wocc} leads to much greater deviations in some cases.
We therefore see that using \eqref{eqn:fbar_inh} yields good (relative to LSDA) performance on ground states; and use Eq.~\eqref{eqn:fbar_inh} for our inhomogeneous effective occupation factor in all subsequent calculations.

Technical details for all atomic and molecular calculations for ground and excited states are in \SMSec{II}.
For now it suffices to say that we carry out LSDA and time-dependent LDA (TDLDA) calculations using standard self-consistent field (SCF) approaches implemented in {\tt psi4}~\cite{Parrish2017,Smith2018}
and {\tt pyscf},~\cite{Sun2017,Sun2020} but evaluate eLDA calculations using an orbital optimized approach with psi4 as an `engine'.
Spin and spatial symmetries are preserved in eLDA calculations, except for atoms which are evaluated
using cylindrical spatial symmetries for consistency with standard quantum chemistry codes and practice.

\subsection{Low-lying excitations in molecules}

\begin{figure}
   \includegraphics[width=\linewidth]{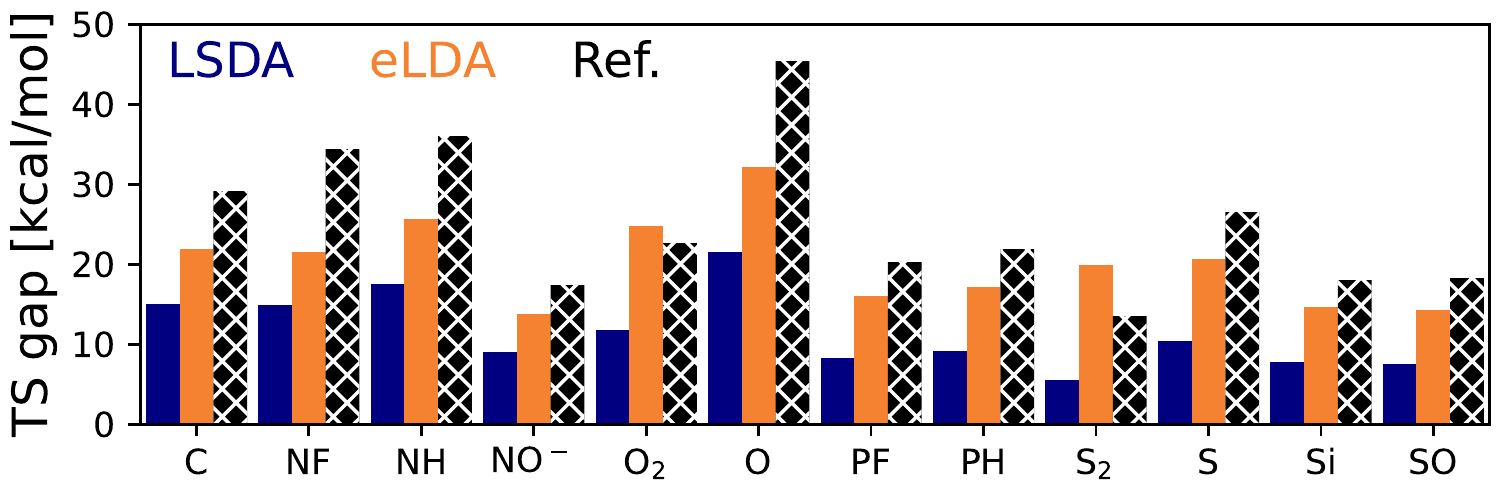} 
   \caption{
   Triplet--singlet gaps in atoms and diatomic systems from
   LSDA (navy) and eLDA (orange) calculations, compared
   to experimental reference data (black with crosses).
   LSDA and reference data from \rcite{Lee2019-Cmplx}.
   TDLDA gaps are too large for the figure, so have been left out.
   \label{fig:TS12}
   }   
\end{figure}

With the eLDA established and validated on ground state systems,
we are ready to test its predictive ability for excitations.
As a first test, (Figure~\ref{fig:TS12}) we consider
the twelve triplet-singlet gaps in biradicals of the
TS12~\cite{Lee2019-TS12} dataset.
The performance of PW92~\cite{PW92} energy differences
(referred to as $\Delta$SCF calculations, to differentiate from TDLDA calculations)
on this dataset was explored in \rcite{Lee2019-Cmplx},
using restricted, unrestricted and complex orbital Kohn-Sham theory.
The mean-signed errors (root mean squared errors) from
$\Delta$SCF calculations are
$-13.7$ ($14.5$) kcal/mol using LSDA (i.e. unrestricted Kohn-Sham
theory); and $10.9$ ($11.5$) kcal/mol for restricted theory.
Employing complex orbitals reduces these LDA errors
substantially, to $-1.2$ ($2.2$) kcal/mol, albeit at the
expense of non-idempotent density matrices.

Using the eLDA formalism developed here (also a $\Delta$SCF method) to compute the gaps yields errors of $-5.0$ ($7.4$) kcal/mol -- respectable statistics and a major improvement on LSDA, as shown in Figure~\ref{fig:TS12}.
Indeed, eLDA is closer in quality to the complex orbital performance than LDA or LSDA performance, despite \TG{}{eLDA being a heavily constrained} `restricted' theory that preserves idempotency (unlike complex orbitals) and avoids spin-contamination issues (unlike unrestricted KS).
\TG{}{This is in partial contrast to the common~\cite{Perdew2021} expectation that symmetry breaking helps DFAs to capture difficult physics.
TDLDA (using VWN correlation~\cite{VWN} and starting from the triplet ground states for consistency with other results) yields enormous errors of 77.2 (88.6) kcal/mol -- too large to include in the figure.}
eLDA thus out-performs both ground state (LSDA) and excited state (TDLDA) LDA-based calculations.

\begin{figure}
   \includegraphics[width=\linewidth]{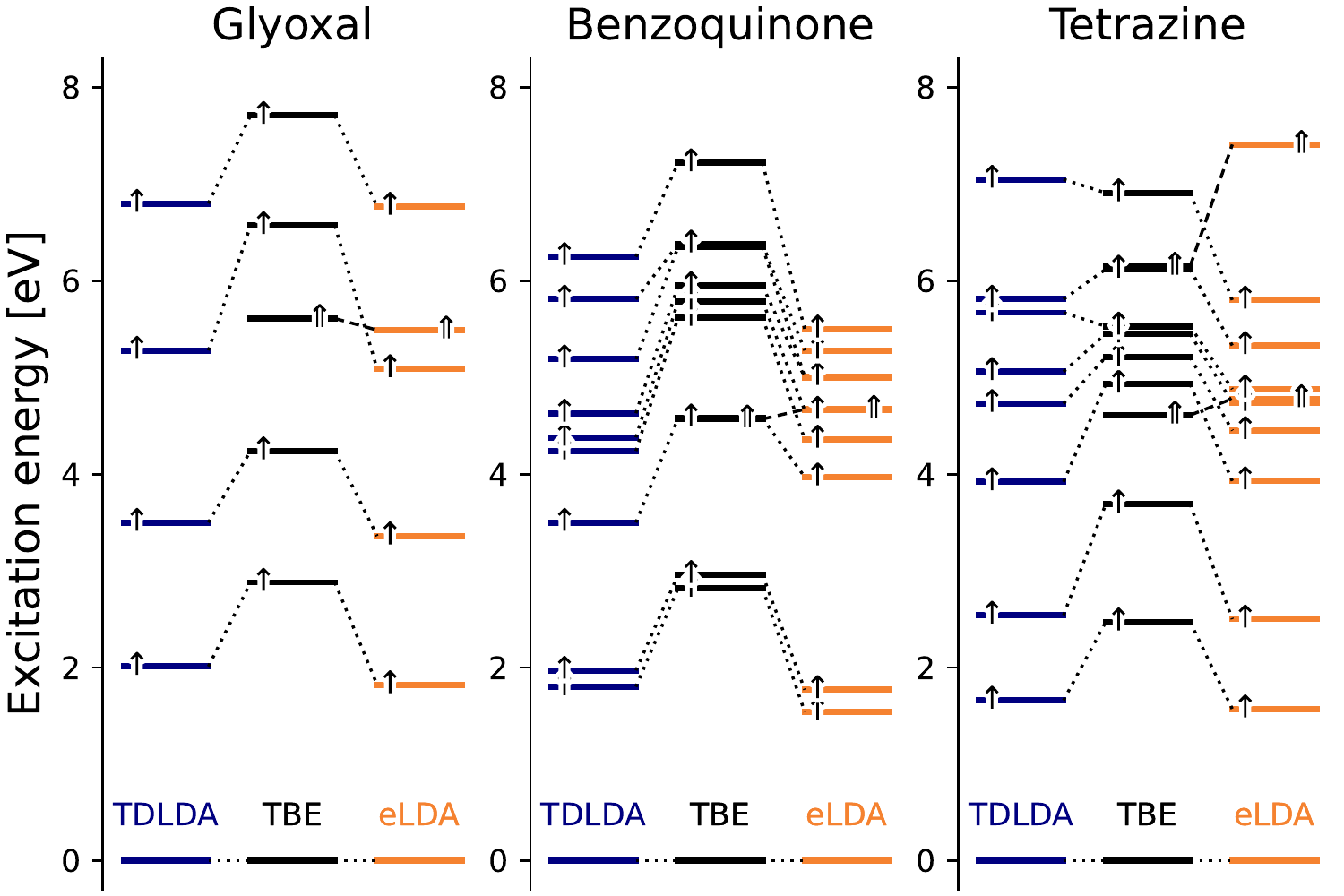} 
   \caption{
   Low-lying spectra (singlets only) of glyoxal, benzoquinone and tetrazine
   predicted using TDLDA (navy) and eLDA (orange);
   compared against theoretical best estimate (TBE) values.~\cite{QuestDB}
   Connections between spectrum in approximations and TBE are shown
   using dotted lines, to facilitate comparisons.
   TDLDA captures single excitations (indicated by single arrows on the level
   line) but misses the double excitations (double arrows) entirely so these
   connections are excluded from the plot.
   \label{fig:Molecules}
   }   
\end{figure}

Continuing on the theme of predicting difficult excitations,
let us consider some excitations that TDLDA cannot predict
at all: double excitations.
Double excitations are singlet excited states in which the
interacting wave function is dominated by a Slater determinant
with paired orbitals, and in which one pair is `doubly promoted'
from the dominant ground state Slater determinant
(e.g. $\iket{\phi_0^2\phi_1^2\phi_3^2}$
instead of $\iket{\phi_0^2\phi_1^2\phi_2^2}$ for a six-electron system).
They are \emph{impossible} to predict using the adiabatic approximation
that is employed in all practical implementations of
time-dependent DFT.~\cite{Maitra2004-Double,Maitra2005}

Figure~\ref{fig:Molecules} shows the low-lying singlet spectra of some selected molecules, computed using adiabatic time-dependent LDA (TDLDA) and eLDA.
We choose glyoxal, benzoquinone and tetrazine from the QuestDB dataset,~\cite{QuestDB} as their low-lying spectra includes difficult-to-predict double excitations for which high-quality theoretical best estimates (TBE) results are available.
They therefore serve as good examples to compare the eLDA approach with its TDLDA counterpart.

It is immediately clear that, for the lowest-lying excitations involving single promotion of an electron (``single excitations'', single arrows), eLDA predicts similar excitation energies to TDLDA and thus has similar performance -- albeit with a slight tendency to underestimate relative to TDLDA.
However, unlike TDLDA, eLDA is also able to predict excitations involving \emph{double} promotion of electrons (``double excitations'', double arrows) with a perfomance similar to that of single excitations.
Thus, eLDA is nearly as good as TDLDA for low-lying excitations involving single promotion of an electron, but is also able to predict double promotions, unlike TDLDA.
It therefore offers a major advance on TDLDA.

\TG{}{We also compute, but do not show, $\Delta$SCF results for excited states, using the maximum overlap method~\cite{Gilbert2008} (MOM) to converge single and double promoted excited states within a self-consistent field framework based on LSDA.
For the MOM calculations we employ the occupation factors from eLDA and promote only the $\up$-electron in single excitations. We use PW92 correlation~\cite{PW92} as it is most similar to our cofe parametrisation.
Results for single excitations are almost identical to eLDA, with an overall MAD of 1.08~eV from MOM compared to 1.05~eV for all single excitations shown in the figure.
However, double excitations are greatly improved by eLDA, with an MAD of 0.67~eV using MOM being reduced to just 0.41~eV for eLDA.
Again, we see that eLDA offers significant improvements.}

Figure~\ref{fig:Molecules} also provides evidence that eLDA can be a cornerstone theory for better excited state approximations, based on the following argument.
As can be seen from the figure, TDLDA and eLDA yield very similar energies for most single excitations.
The similarity of TDLDA and eLDA energies suggests that \emph{all regular DFAs} are likely to yield similar energies for these excitations, whether evaluated as TDDFAs or eDFAs -- a theoretical justification for this argument is provided in \SMSec{III}.
Thus, the thirty years of refinement of generalized gradient approximations (GGAs) and meta-GGAs (MGGAs) that has improved the quality of spectra predicted using TD(M)GGAs is likely to similarly improve spectra evaluated using e(M)GGAs.
But e(M)GGAs may \emph{also} exploit the extra degree of freedom enabled by the use of cofe-gas physics and effective $\fb(\vr)$.

In summary, we see that TDLDA fails quite dramatically for TS12 (Figure~\ref{fig:TS12} and related discussion) and cannot capture double excitations (Figure.~\ref{fig:Molecules}); in contrast to an excellent (TS12) or impressive (double excitations) performance from eLDA on the difficult excitations.
Errors in single excitation spectra (Figure.~\ref{fig:Molecules}) from TDLDA and eLDA are similar.
Directly, this shows that eLDA either improves excited state predictions, or does not make them worse.
Indirectly, it has positive implications for refinements to eLDAs, e.g. eGGAs or eMGGAs.
\\

\section{Future prospects and conclusions}
\label{sec:Conclusions}

Ensemble density functional theory has recently benefited from a surge of fundamental understanding.
This has led to rapid advancements in extending, to excited states, the power of density functional theory for computing electronic structure of ground states.
Especially, EDFT deals seamlessly with highly ``quantum'' states~\cite{Gould2021-DoubleX}
(e.g. superpositions of Slater determinants and double excitations)
of relevance to solar energy applications and quantum technologies.

However, despite an accumulation of successful applications, EDFT currently lacks a fully consistent framework for improving approximations: in the sense that it borrows density functional approximations (DFAs) which were originally designed for ground states as the key building blocks of the extended DFAs for excited states.
\TG{}{Especially, computationally favorable modelling of singlet-singlet excitations does not include all the relevant correlations.~\cite{deSilva2019}}

This work takes a first step toward \emph{deriving a novel family of DFAs specifically designed for excitations}.
It presents \TG{}{(Sections~\ref{sec:cof} and \ref{sec:Inhomogeneous})} the cornerstone models: the `cofe' homogeneous electron gas (HEG) and the LDA for EDFT (eLDA).
The `cofe' HEG is developed using an unnoticed  -- thus, so far, unexplored -- class of non-thermal ensemble states of the the HEG.
Analytic expressions of the relevant (defined by two parameters, like LSDA) energy components are reported in Table~\ref{tab:HEGs}.
Some of these components have no analogues in regular DFT but find home and use in EDFT.
High- and low-density limits of the correlation energy have been found analytically.

The eLDA is derived by dividing the DFT energy expression into terms that need to be treated using the inhomogeneous system, and those that are locally approximated using a cofe gas.
Parametrisations for all terms required by the eLDA are derived and provided.
\TG{}{An ansatz (consistent with one-electron, unpolarized and cofe gas limits) is also made for the effective occupation factor of inhomogeneous systems, and is `normed' on doublet systems in lieu of semi-classical results.}

The novel eLDA is then tested on a suite of important examples including ionization potentials, small triplet-singlet gaps, and low-lying excitations.
These examples reveal that eLDA performs similarly to LSDA and/or time-dependent LDA (TDLDA) on problems where standard theories are known to work.
However, it \emph{also performs very effectively on problems where LSDA/TDLDA fail} -- yielding excellent triplet-singlet gaps and impressive double excitation energies.

eLDA therefore readily offers an effective alternative to standard polarized-gas based theories for both ground and excited state problems.
But, we stress that its true potential lies as the cornerstone for {\em better} models and methodologies.
\emph{What are the next natural steps to be considered?}
We finish with three suggestions.

(I) It is vital to develop a generalized gradient approximation (GGA) for cofe-HEGs, to yield an eGGA along the lines of Eq.~\eqref{eqn:eLDA}.
The development of accurate GGAs in the late-1980s/early-90s greatly accelerated interest in DFT for ground states, by giving answers that were usefully predictive.
eGGAs should do the same for excited states.
Importantly, eGGAs would seamlessly integrate with existing hybrid-EDFT successes~\cite{Gould2020-Molecules,Gould2021-DoubleX,Gould2022-pEDFT} and remove reliance on combination rules that (despite working unexpectedly well) are known to be incorrect for correlation.~\cite{Gould2021-DoubleX}
From there, additional steps may readily be taken up an \emph{excited state} Jacob's ladder,~\cite{Perdew2001-Jacob} to gain systematic improvements in excited state DFT modelling.

(II) The optimal way to model $\fb(\vr)$ remains an open problem, and is entangled with (I) \TG{}{and semi-classical arguments}.
It would be useful to understand why Eq.~\eqref{eqn:fbar_inh} works so much better than Eq.~\eqref{eqn:fbar_wocc} \TG{}{and how it can be improved}.
Exploiting exact relationships, like combination rules~\cite{Gould2021-DoubleX}, is likely to lead to improved understanding and adaptation of $\fb(\vr)$ in inhomogeneous systems; and thus improvements to the predictive ability of eLDA and any eDFAs built on it.
\TG{}{Improved understanding of finite yet uniform electron gases may also help.~\cite{Loos2012,Loos2013-Ringium,Loos2014-Ringium,Loos2014}}

(III) The cofe-gas is not the only excited state (ensemble) HEG that we could have used.
As discussed in Section~\ref{sec:cof} it is a logical and simple \emph{two-}parameter model that yields appropriate limits yet incorporates excited state physics \TG{}{in a way that is consistent with known conditions}.
But, allowing for more parameters provides a wide scope for further generalizations \TG{}{and improvements}.
For example, \TG{}{in the spirit of Samal and Harbola,~\cite{Samal2005}} one might separate the density into core (density, $n_{\text{core}}$) orbitals that are all double occupied, and use a cofe-like treatment for the remaining orbitals -- yielding a three-parameter HEG governed by $n$, $n_{\text{core}}/n$ and $\fb$ that includes excited states.

Python code for studying and implementing the theory work in this manuscript is provided on Github~\url{https://github.com/gambort/cofHEG}.
Code to reproduce the atomic and molecular tests is available on request.

\acknowledgments

TG was supported by an Australian Research Council (ARC) Discovery Project (DP200100033) and Future Fellowship (FT210100663).
Computing resources were provided by the National Computing Merit Application Scheme (NCMAS sp13).
TG and SP would like to thank Paola Gori-Giorgi for interesting discussions regarding homogeneous electron gases and their low density limit; useful discussions with Marco Govoni on a previous version of the manuscript are also acknowledged.

\appendix

\section{Exchange properties of {cofe HEGs}}
\label{app:Functionals}

Both $\epsx$ and $\Pi_{\xrm}$ involve integrals of form,
\begin{align}
X_2[f]:=&\int_0^{\infty}\int_0^{\infty}f_{\max(q,q')}A(q,q')
\frac{{q'}^2dq'}{2\pi^2}\frac{q^2dq}{2\pi^2}
\nonumber\\
=&2\int_0^{\infty}\int_{q'}^{\infty}f_{q}A(q,q')
\frac{q^2dq}{2\pi^2}\frac{{q'}^2dq'}{2\pi^2}
\nonumber\\
=&2\int_0^{\infty}\int_0^{\infty}
\Theta(q-q')f_{q}A(q,q')
\frac{{q'}^2dq'}{2\pi^2}\frac{q^2dq}{2\pi^2}
\nonumber\\
=&\int_0^{\infty}f_q \bar{A}(q)\frac{q^2dq}{2\pi^2}
\label{eqn:X2}
\end{align}
where $A(q,q')=A(q',q)$ and,
\begin{align*}
\bar{A}(q)=&2\int_0^{q} A(q,q') \frac{{q'}^2dq'}{2\pi^2}
\end{align*}
For $\epsx$ we have $A(q,q')=V(q,q')$ where
$\bar{V}(q)=2\int_0^q V(q,q')\tfrac{{q'}^2dq'}{2\pi^2}
=\tfrac{q}{\pi}\int_0^1 \log\tfrac{|1+x|}{|1-x|}xdx
=\tfrac{q}{\pi}$. We thus obtain
eq.~\eqref{eqn:epsxfq} of the main text.
To compute $\Pi_{\xrm}$ we can set $A(q,q')=1$
where $\bar{1}(q)=2\int_0^q\tfrac{{q'}^2dq'}{2\pi^2}
=\tfrac{q^3}{3\pi^2}$ and so we can
easily compute $\Pi_{\xrm}$ given $f_q$.
For cofe HEGs we obtain
eqs~\eqref{eqn:epsxcof} and \eqref{eqn:Pixcof}.

The case of $n_{2,\xrm}(R)$ is also covered by
\eqref{eqn:X2}, by setting
$A(q,q';R)=\int e^{i(\vq-\vqp)\cdot\vR}
\tfrac{d\vech{q}}{4\pi}\tfrac{d\vech{q}'}{4\pi}$.
However, this is rather painful to deal with
in general. The special case of cofe HEGs
is more easily handled by recognising that
$f_{\max(q,q')}=\fb \Theta(\kFb-q)\Theta(\kFb-q')$.
Then,
\begin{align}
n_{2,\xrm}^{\cof}(R)=&-\fb
\int_0^{\kFb} e^{i\vq\cdot\vR} \frac{d\vq}{(2\pi)^3}
\int_0^{\kFb} e^{-i\vqp\cdot\vR} \frac{d\vqp}{(2\pi)^3}
\\
=&-\fb \bigg|\frac{\kFb^3}{6\pi^2}
g(\kFb R)\bigg|^2\equiv -\Pi_{\xrm}^{\cof} N(\kFb R)
\label{eqn:gxapp}
\end{align}
where $\Pi_{\xrm}=-\frac{n^2}{\fb}$,
$g(x)=3[\sin(x)-x\cos(x)]/x^3$ and
$N(x)=|g(x)|^2$. Thus, we obtain
eq.~\eqref{eqn:n2xcof}.

\section{Hartree properties of cofe HEGs}
\label{app:Hartree}

Let us consider eqs~\eqref{eqn:Ewn2} and \eqref{eqn:n2H}
for the special case of an HEG. First, we note that,
$n_{\FE\FE}=n$ for every state and therefore,
$n_{2,\Hrm}=n^2 + \Delta n_{2,\Hrm}$ where
$\Delta n_{2,\Hrm}(\vr,\vrp)
=\sum_{\FE\neq \FE'}w_{\max(\FE,\FE')}
n_{\FE\FE'}(\vr)n_{\FE'\FE}(\vrp)$. Furthermore,
the resulting pair-density can depend only on
$\vR=\vr-\vrp$ while symmetry means it depends only
on $R=|\vr-\vrp|$. Thus,
\begin{align}
\depsH = \frac{1}{N}\Delta E_{\Hrm}
=\frac{1}{n} \int \Delta n_{2,\Hrm}(R) \frac{4\pi R^2 dR}{2R}
\label{eqn:depsHapp}
\end{align}
where we used $n^2$ to cancel the
background charge, $N=n V$ to cancel the integral
over $\vr$, and symmetry to simplify the remaining
integral over $\vrp=\vr+\vR$.
Our goal is therefore to determine $\Delta n_{2,\Hrm}(R)$,
Note, the working in this appendix is rather involved,
so we will often drop superscripts $^{\wv}$ in working.

We are now ready to look at HEG ensembles.
Consider a finite HEG of $N$ electrons in a volume $V$,
with density $n=N/V$. The orbitals are
$\phi_{\vq} \approx \tfrac{1}{\sqrt{V}} e^{i\vq\cdot\vr}$
for $\vq$ on an appropriate reciprocal space grid.
Each state, $\iket{\FE}$ has density
$n_{s,\FE\FE}(\vr)=N/V=n$. 
The ground state is $\iket{0}=\iket{\vq_1^2\cdots \vq_{N/2}^2}$
and is unpolarized.
Other states may be described using
$\iket{\FE}=\hat{P}_{\vQ_\FE}\iket{0}$
where $\hat{P}$ promotes Fock orbitals
in the Slater determinant and
$\vQ_{\FE}:=_{\vq_{i_1}\cdots \vq_{i_p}}^{\vq_{a_1}\cdots \vq_{a_p}}$
contains lists of from ($i\leq N/2$) and to ($a>N/2$) orbitals,
including spin. Cross-densities, when $\FE\neq \FE'$, are
$n_{s,\FE\FE'}(\vr)=e^{i\Delta\vq_{\FE\FE'}\cdot\vr}/V$
or zero. The former result occurs if and only if
$\vQ_{\FE}$ and $\vQ_{\FE'}$ differ by a single orbital of
the same spin, giving $\Delta\vq_{\FE\FE'}=\vq_{\in \FE}-\vq_{\in \FE'}$.
We use ``connected'' (con) to refer to any pair
of states $\FE$ and $\FE'$ that differ only by a single
orbital, and call $\Delta\vq_{\FE\FE'}$ the connection
wavenumber.

\newcommand{\NT}{{\cal N}_T}

Let us now consider the case that $N$ electrons are
assigned to $N/2\leq M\leq N$ orbitals, for a mean
occupation of $f=\tfrac{N}{M}$. There are
$\NT=\tfrac{(2M)!}{N!(2M-N)!}$ total states once spin is
accounted for, each of which is weighted by,
$w=\tfrac{1}{\NT}$.
Each of the $\NT$ states, $\iket{\FE}$,
has $N_{\up,\down,\FE}$ electrons of each spin, 
giving $\zeta_{\FE}=\tfrac{N_{\up,\FE}-N_{\down,\FE}}{N}$.
State $\iket{\FE}$ is connected to $C_{\FE}$ other
states. Since only one orbital may change at a time,
we obtain $C_{\FE}=N_{\up,\FE}(M-N_{\up,\FE})
+ N_{\down,\FE}(M-N_{\down,\FE})
=NM-\tfrac{N^2}{2}(1+\zeta_{\FE}^2)$,
where $N_{\up,\down,\FE}\leq M$.

Our goal is to obtain useful properties of the Hartree
pair-density. The pair-density is defined by,
\begin{align}
\Delta n_{2,\Hrm}(\vR)=
\frac{1}{\NT}\sum_{\FE,\FE'~\text{con}~\FE}
\frac{e^{i\Delta\vq_{\FE\FE'}\cdot\vR}}{V^2}\;,
\label{eqn:n2Hcomb}
\end{align}
where we used $w_{\FE}=w_{\FE'}=\tfrac{1}{\NT}$
and $\vr-\vrp:=\vR$.
The special case of $\vr=\vrp$ ($\vR=\vec{0}$)
yields the ``on-top'' pair density deviation,
$\Pi_{\Hrm}=\Delta n_{2\Hrm}(\vR=0)$,
which is relatively straightforward to evaluate
using, $\Delta n_{2,\Hrm}=
\frac{1}{\NT V^2}\sum_{\FE}C_{\FE}$,
which follows from $e^{i\Delta\vq\cdot\vR}=1$ for all
connected states, and the definition of $C_{\FE}$.
Using $C_{\FE}$ from the above paragraph yields,
\begin{align}
\Delta \Pi_{\Hrm}=&
\tfrac{1}{V^2}
[NM+\tfrac{N^2}{2}(1+\bar{\zeta}^2)]
=n^2\big[\tfrac{1}{f}-\tfrac{1+\bar{\zeta}^2}{2}
\big]
\;.
\label{eqn:PiHcomb}
\end{align}
where $\bar{\zeta}^2=\tfrac{1}{\NT}\sum_{\FE}\zeta_{\FE}^2$
is the ensemble averaged of the squared spin-polarization.

As an initial test, consider the above analysis for
the two special types of gases, unpolarized and fully
polarized gases, which have no ensemble effects 
and which must therefore yield $\Delta \Pi_{\Hrm}=0$.
An unpolarized gas involves
$M=N/2$, $\fb=2$, $\NT=1$ and $\zeta_{\FE}=0$, yielding
$\Delta \Pi_{\Hrm}=n^2(\tfrac{1}{2}-\tfrac{1}{2})=0$.
A fully polarized gas involves
$M=N$, $\fb=1$, $\NT=1$ and $\zeta_{\FE}=1$, yielding
$\Delta \Pi_{\Hrm}=n^2(\tfrac{1}{1}-1)=0$.
Thus, both exhibit the expected behaviour.
We therefore see that eq.~\eqref{eqn:PiHcomb}
is consistent with existing results.

We are now ready to generalize to
constant occupation factor (cof) gases,
with $f_q = \fb\Theta(\kFb-q)$,
for $1<\fb<2$. As discussed in the main text,
we restricted to the special case of
maximally polarized states, $\iket{\FE}$,
in which each state has the maximum
spin-polarization allowed by $\fb$.
All these states involve
$N_{\up}=M$ and $N_{\down}=N-M$
giving, $\zeta_{\FE}=\tfrac{2M-N}{N}
=\tfrac{2}{\fb}-1=\bar{\zeta}$.
Eq.~\eqref{eqn:PiHcomb} then yields,
\begin{align}
\Delta \Pi_{\Hrm}=& n^2\frac{(2-\fb)(\fb-1)}{\fb^2}\;,
&
\Pi_{\Hrm}=& n^2\frac{3\fb-2}{\fb^2}\;.
\label{eqn:PiHinApp}
\end{align}
for the on-top, $\vR=\vec{0}$, pair-density.

We are now ready to move on from the on-top
hole to consider general $\vR\neq \vec{0}$.
We first recognise that equal waiting of states
is equivalent to equal weighting of connection
wavenumbers, yielding,
\begin{align}
\Delta n_{2,\Hrm}(\vR)
=&\Delta \Pi_{\Hrm}
\frac{1}{M^2}\sum_{\vq}\sum_{\vq'\neq \vq}
e^{i(\vq-\vq')\cdot\vR}
\nonumber\\
=&\Delta \Pi_{\Hrm}\big[ |g(\kFb R)|^2 - \tfrac{1}{M}\big]
\;,
\end{align}
where $g:=\tfrac{1}{M}\sum_{\vq}e^{i\vq\cdot\vR}$.
We next impose symmetry on the wavenumbers, and
approximate the sum by an integral to obtain,
\begin{align}
g(\kFb R)\approx&\frac{1}{M}\int_0^{(\tfrac{3M}{4\pi})^{1/3}}
\frac{\sin(qk_VR)}{qk_VR} 4\pi q^2 dq
\nonumber\\
=&\frac{3[\sin(\kFb R)-\kFb R\cos(\kFb R)]}{(\kFb R)^3}
\end{align}
where $k_V=2\pi/V^{1/3}$ is the wavenumber associated
with the volume $V$; and
$\kFb=(6\pi^2 M/V)^{1/3}=(6\pi^2 n/\fb)^{1/3}$
is the usual Fermi wavenumber.
$g(x)$ the same expression found in eq.~\eqref{eqn:gxapp}.
Note, $\Delta n_{2,H}(R)$ integrates to zero, as expected.

Finally, eq.~\eqref{eqn:depsHapp} becomes
$\depsH=\tfrac{\Delta \Pi_{\Hrm}}{n}\allowbreak
[ \int_0^{\infty} \tfrac{g(\kFb R)^2}{R}\allowbreak 2\pi R^2dR
- \tfrac{\fb}{2n}(\tfrac{9\pi}{2V})^{1/3} ]$.
In the limit $V\to\infty$ the second term vanishes,
yielding,
\begin{align}
\depsH^{\cof}
=&\fb\Delta \Pi_{\Hrm}|\epsx^{\cof}(r_s,\fb)|
=\frac{C_{\Hrm}}{r_s}\frac{(2-\fb)(\fb-1)}{\fb^{4/3}}
\\
=&|\epsx^{\cof}(r_s,\fb)|\frac{(2-\fb)(\fb-1)}{\fb}
\label{eqn:depsHinApp}
\end{align}
where we used, $\epsx^{\cof}=-\frac{\fb}{n}\int_0^{\infty}
\frac{g(\kFb R)^2}{R}2\pi R^2dR$ [which follows from
$n_{2,\xrm}(R)=-\fb g(\kFb R)^2$]
and $\epsx^{\cof}=\tfrac{-C_{\xrm}}{r_s}[2/\fb]^{1/3}$
derived in the main text,
to obtain $C_{\Hrm}=2^{1/3}C_{\xrm}=0.577252$.
Similarly,
\begin{align}
n_{2,\Hrm}(R)=n^2 + \Delta n_{2,\Hrm}(R)
=&n^2+\Delta \Pi_{\Hrm}^{\cof}N(\kFb R)
\end{align}
where $\Delta \Pi_{\Hrm}^{\cof}$ is defined
in eq.~\eqref{eqn:PiHinApp}; and
$N(x)=g(x)^2=9[\sin(x)-x\cos(x)]^2/x^6$
is the unitless function defined near
eq.~\eqref{eqn:n2xcof} or \eqref{eqn:gxapp}.

\section{State-driven correlation of cofe HEGs}
\label{app:CorrAll}

\subsection{State-driven correlation energy from
the random-phase approximation}
\label{app:CorrRPA}

\begin{figure}
\includegraphics[width=\linewidth]{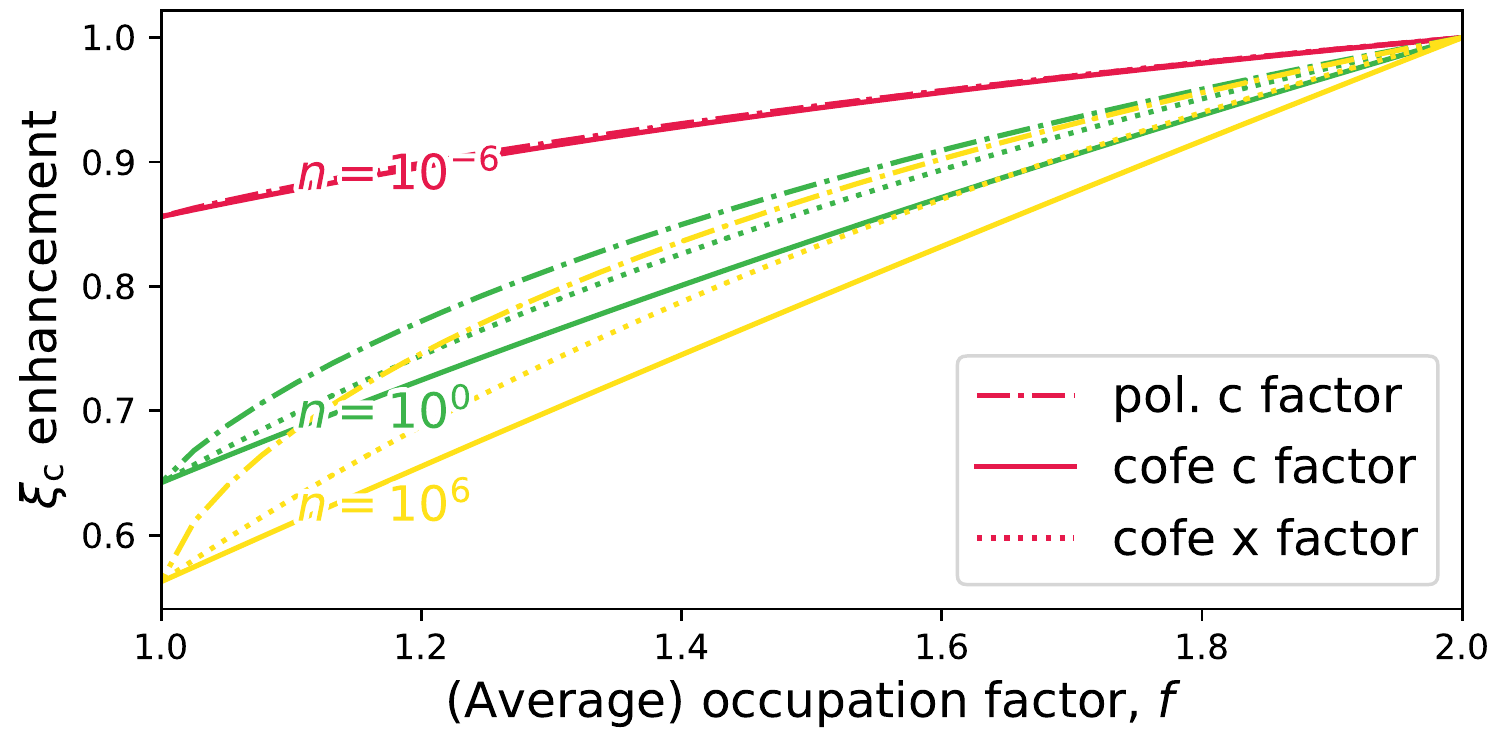} 
\caption{Correlation enhancement of spin-polarized (dash-dot lines)
and cofe (solid lines) HEGs as a function of occupation
factor, $f$. The cofe ot exchange factor (dotted lines) is also shown
after rescaling to yield the same vales for $\fb=1$ and $\fb=2$.
\label{fig:RPA}
}
\end{figure}

The state-driven correlation energy [eq.~\eqref{eqn:epscFDT}]
involves the response function at imaginary frequencies.
The imaginary frequency density-density response
of an unpolarized HEG is,
\begin{align}
\chi_0(q,i\omega;r_s):=-\frac{k_F}{4\pi^2}C(\tfrac{q}{2k_F},\tfrac{\omega}{qk_F})
\end{align}
where $k_F=1.9191583/r_s=(3\pi^2n)^{1/3}$ is the
Fermi wavenumber of an unpolarized gas. Here,
\begin{align}
C(Q,\Gamma)=&1 + \frac{\Gamma^2-Q_+Q_-}{4Q}
\log\frac{Q_+^2+\Gamma^2}{Q_-^2+\Gamma^2}
\nonumber\\&
+ \Gamma\bigg[\tan^{-1}\frac{Q_-}{\Gamma}-\tan^{-1}\frac{Q_+}{\Gamma}\bigg]
\label{eqn:CLind}
\end{align}
where $Q_{\pm}=Q\pm 1$.
For brevity we shall use $Q=\tfrac{q}{2k_F}$ and
$\Gamma=\tfrac{\omega}{qk_F}$ to always mean
unpolarized gas quantities.

For a polarized HEG we take half of two unpolarized
systems with
$k_{F\up}=k_F(1+\zeta)^{1/3}:=k_F h_+$
and $k_{F\down}=k_F(1-\zeta)^{1/3}:=k_F h_-$.
Therefore,
\begin{align}
\chi_0(q,\omega;r_s,\zeta)=&
-\tfrac{k_F}{8\pi^2}
\big[ h_+
C(\tfrac{Q}{h_+},\tfrac{\Gamma}{h_+})
+ h_-
C(\tfrac{Q}{h_-},\tfrac{\Gamma}{h_-})
\big]
\label{eqn:chi0pol}
\end{align}
The cofe case of constant $\fb$ is
easily dealt with by including a prefactor
of $\fb$ on $\chi_0$, and using the cofe Fermi level,
$\kFb= k_F(2/\fb)^{1/3}:=k_F g$.
It follows from $\fb=2/g^3$ that,
\begin{align}
&\chi_0^{\cof}(q,\omega;r_s,\fb)=
-\tfrac{k_F}{4\pi^2g^2}
C(\tfrac{Q}{g},\tfrac{\Gamma}{g})\;.
\label{eqn:chi0cof}
\end{align}
Setting $\zeta=0$ and $\fb=2$ yields $h_{\pm}=g=1$
and yields the same response as the unpolarized gas.
Similarly, setting $\zeta=1$ in \eqref{eqn:chi0pol}
gives the same result as setting $\fb=1$ in \eqref{eqn:chi0cof},
as expected.

From the response function we are able to evaluate the
random-phase approximation for the correlation energy,
via,
\begin{align}
\epsc^{\RPA}
=&\tfrac{1}{2n}\int_0^{\infty}\tfrac{d\omega}{\pi}
\int_0^{\infty}\tfrac{q^2dq}{2\pi^2}
[\chi_0\tfrac{4\pi}{q^2}+\log(1-\chi_0\tfrac{4\pi}{q^2})]\;.
\end{align}
This may be made more convenient by using
$n=\tfrac{k_F^3}{3\pi^2}$,
$q=2k_F Q$ and $\omega=qk_F\Gamma$
to write,
\begin{align}
\epsc^{\RPA}
=&\frac{12k_F^2}{\pi}
\II{\tfrac{\pi}{k_F^2 Q^2}\chi_0}\;.
\end{align}
where $\II{f}:=\int_0^{\infty}d\Gamma \int_0^{\infty}Q^3dQ
[-f+\log(1+f)]$. We may also define,
$\IB{P} = \II{\tfrac{P}{Q^2}C(Q,\Gamma)}$.

Thus, the RPA enhancement factor for a polarized,
relative to an unpolarized gas at the same density, gas is,
\begin{align}
\xi_{\crm}^{\RPA}(\zeta)
=&\frac{\II{ \tfrac{P h_+}{2Q^2}C(\tfrac{Q}{h_+},\tfrac{\Gamma}{h_+})
+ \tfrac{P h_-}{2Q^2}C(\tfrac{Q}{h_-},\tfrac{\Gamma}{h_-}) }}%
{\II{\tfrac{P}{Q^2}C(Q,\Gamma) }}
\nonumber\\
=&\frac{h_+^5\IB{\tfrac{P}{2h_+}} + h_-^5\IB{\tfrac{P}{2h_-}}}{\IB{P}}
\end{align}
where $h_{\pm}=(1\pm\zeta)^{1/3}$.
The equivalent enhancement factor of a cof
ensemble HEG may be written as,
\begin{align}
\xi_{\crm}^{\RPA,\cof}(\fb)
=&\frac{\II{ \tfrac{P}{g^2Q^2} C(\tfrac{Q}{g},\tfrac{\Gamma}{g}) }}%
{\II{\tfrac{P}{Q^2}C(Q,\Gamma) }}
=\frac{g^5\IB{\tfrac{P}{g^4}}}{\IB{P}}
\end{align}
where $P:=\tfrac{1}{2\pi k_F}=0.08293 r_s$ and
$g=(2/\fb)^{1/3}$.

The RPA enhancement is expected to be accurate
in the high-density of matter $k_F\to\infty$.
Figure~\ref{fig:RPA} shows (state-driven)
correlation energy enhancement factors,
$\xi_{\crm}^{\RPA}(\zeta)$ and $\xi_{\crm}^{\RPA,\cof}(f)$
as a function of the (average) occupation factor, $\fb$,
using $\zeta=\fmap^{-1}(\fb)$
[from eq.~\eqref{eqn:fmapInv}]
for the effective spin-polarization.
It reports $\xi$ for
high ($n=10^6$), medium ($n=1$) and low
($n=10^{-6}$) densities.
We see that the state-driven correlation energy of
cofe HEGs is:
i) virtually linear in $\fb$, for high densities;
ii) very similar to the (renormalized) on-top exchange
enhancement factor, for low densities.

The high density ($r_s\to 0$) behaviour of $\xi_{\crm}^{\cof}$ 
can be shown analytically, because $P\to 0$.
We may therefore Taylor expand the log to obtain,
\begin{align}
\lim_{P\to\infty}\IB{P}\approx&
\int_0^{\infty}Q^3 dQ\int_0^{\infty}d\Gamma	
\tfrac12\big(\tfrac{P}{Q^2}C\big)^2\;,
\end{align}
from which it follows that $\xi_{\crm}^{\cof}
=g^5\IB{P/g^4}/\IB{P}
=g^5(\tfrac{1}{g^4})^2=g^{-3}=\fb/2$ is linear in $\fb$.
The RPA is not appropriate for the low density limit,
although we shall later see it is qualitatively correct.

\newcommand{\DEH}{\Delta E_{\Hrm}^{\wv,0}}
\newcommand{\DEHl}{\Delta E_{\Hrm}^{\wv,\lambda}}
\newcommand{\DEHi}{\Delta E_{\Hrm}^{\wv,\infty}}

\subsection{State-driven correlation energies in general}
\label{app:CorrGen}

We are now ready to use what we have learned
about correlation energies from the RPA and
theoretical arguments to obtain general
expressions for the SD correlation energies
in cofe HEGs.
Let us begin with the low-density limit. The main text
has shown that,
\begin{align}
\lim_{r_s\to\infty}\epsc^{\cof}(r_s,\fb) =&
\epsx(r_s)\big[\tfrac{C_{\infty}}{C_{\xrm}} - f_{\Hx}^{\cof}(\fb)\big]\;.
\end{align}
It can also be shown that $\lim_{r_s\to\infty}\epsc^{\DD,\cof}(r_s,\fb) \to -\depsH^{\cof}(r_s,\fb)$ -- this result is a specialized case of a broader relationship to be discussed in a future work.
It is thus clear that the SD enhancement factor must capture the low-density scaling and cancel exchange:
\begin{align}
\lim_{r_s\to\infty}\epsc^{\SD,\cof}(r_s,\fb) \to&
\epsx(r_s)\big[\tfrac{C_{\infty}}{C_{\xrm}} - f_{\xrm}^{\cof}(\fb)\big]\;.
\label{eqn:epscld_app}
\end{align}
Surprisingly, this is consistent with the low-density
behaviour shown in Figure~\ref{fig:RPA} and so
reveals that the RPA is qualitatively
correct even in the low-density limit. 

In the high-density limit, we instead obtain,
\begin{align}
\lim_{r_s\to 0}\epsc^{\SD,\cof}(r_s,\fb) \to &
(\fb-1)\epsc^U + (2-\fb)\epsc^P
\nonumber\\
=&\epsc^U + (2-\fb)[\epsc^P-\epsc^U]
\label{eqn:epschd_app}
\end{align}
where $\epsc^U:=\epsc(r_s,0)$ is the correlation energy of an
unpolarized gas and $\epsc^P:=\epsc(r_s,1)$ is the correlation energy
of a fully polarized gas
and $2-\fb=[\xi^{\RPA,\cof}(\fb)-\xi^{\RPA}(0)]%
/[\xi^{\RPA}(1)-\xi^{\RPA}(0)]$.
This is analogous to the known result that,
\begin{align}
\lim_{r_s\to 0}\epsc(r_s,\zeta) \to&
\epsc^U + H^{\RPA}(\zeta)[\epsc^P-\epsc^U]\;,
\label{eqn:epscWithRPA}
\end{align}
where $H^{\RPA}(\zeta)=-2[I(\zeta)-1]$ is
obtained from eq.~(32) of \rcite{WP91} or, equivalently,
$H^{\RPA}(\zeta) := [\xi^{\RPA}(\zeta)-\xi^{\RPA}(0)]%
/[\xi^{\RPA}(1)-\xi^{\RPA}(0)]$.
We cannot say anything about the DD correlation energy in this limit.

\begin{figure}
\includegraphics[width=\linewidth]{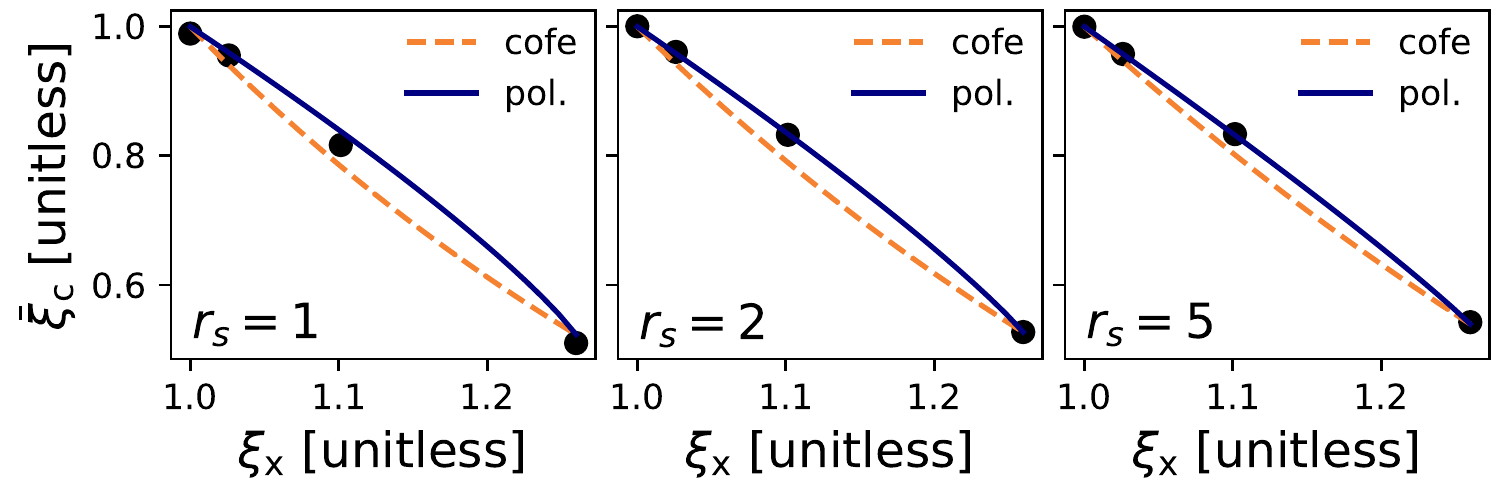} 
\caption{
$\bar{\xi}_{\crm}$ versus $\xi_{\xrm}$ using RPA data
for cofe HEG (orange, dashed lines) and
polarized HEG (navy, solid lines). Black dots
indicate data from \rcite{Spink2013}.
\label{fig:xiMapping}
}
\end{figure}

We thus obtain limiting behaviours for 
high and low-density HEGs. In typical
polarized gases, one uses expansion in
both limits together with QMC data,
$\epsc^{\QMC}(r_s,\zeta)$, to fill in the
gaps for moderate and large densities.
We do not have QMC data for cofe gases. Thus,
the final step in our analysis of correlation
energies is to show how to reuse existing
polarized gas QMC data for cofe HEGs.

As a first step, we assume that the high-density
relationship [eq.~\eqref{eqn:epscWithRPA}] between
RPA and exact results is true for moderate and large
$r_s$. That is, we expect
\begin{align}
\bar{\xi}_{\crm}
:=1 + \frac{\xi_{\crm}(\zeta=1)-1}{\xi_{\crm}^{\RPA}(\zeta=1)-1}
\big[ \xi_{\crm}^{\RPA} - 1 \big]
\approx \xi_{\crm}^{\QMC}
\end{align}
to be approximately valid for all $r_s$. The usefulness
of this approximation is further supported by
Figure~\ref{fig:RPA}, which shows that the
RPA yields an approximately linear dependence
on $\fb$ even for low-density HEGs where
the RPA is expected to be poor.

The second step is to recognise that, in low-density gases,
we may write,
$\xi_{\crm}=\tfrac{C_{\infty}}{C_{\xrm}}-\xi_{\xrm}$ and
$\xi_{\crm}^{\SD,\cof}=\tfrac{C_{\infty}}{C_{\xrm}}+\xi_{\xrm}^{\cof}$
and therefore 
$\epsc^{\SD,\cof}(r_s\to 0,\fb)\approx X(\epsx^{\cof}(r_s,\fb))$
where $\epsc(r_s\to 0,\zeta):=X(\epsx(r_s,\zeta))$
-- here $X$ is a single-variable function.
Figure	~\ref{fig:xiMapping} shows that a similar
result nearly holds for moderate $r_s$ and,
furthermore, that models of both cofe and polarized
gases agree rather well with
QMC data from Spink \emph{et al}~\cite{Spink2013},
despite the data being for polarized gases.
We therefore assume that,
\begin{align}
\epsc^{\SD,\cof}(r_s,\fmap(\zeta))\approx&
\epsc^{\QMC}(r_s, \zeta)
\label{eqn:epscQMC_app}
\end{align}
for moderate and large densities with viable
QMC data, where $\fmap$ is defined such that
$\epsx(\zeta)=\epsx^{\cof}(\fb=\fmap(\zeta))$.
This is a rather good approximation in practice
as the maximum difference between $\epsc^{\SD,\cof}$
using RPA and $\epsc^{\SD,\cof}$ using
\eqref{eqn:epscQMC_app} is 1~mHa for
$r_s=1$, and is sub-mHa for larger $r_s$.

Thus, equations \eqref{eqn:epschd_app}, \eqref{eqn:epscQMC_app} and \eqref{eqn:epscld_app} provide a set of constraints and reference values (from existing QMC data) for high, moderate and low densities, respectively.
These three relationships are used in Appendix~\ref{app:Param} to produce the parametrisation for the state-driven correlation energy of a cofe HEG.

\TG{}{It would be very desirable to obtain QMC or similar-quality reference data for cofe HEGs, to provide direct inputs for parametrizations}.
The derivative, $d e^{\cof}(r_s,\fb)/d\fb|_{\fb=2}$, may be amenable to computation using existing techniques, as it involves only low-lying excited states.

\section{Parameterizations}
\label{app:Param}

The main text and previous appendices have introduced five terms that go into the cofe HEG energy as a function of $r_s$ and $\fb$.
This appendix will provide a useful parametrisation of the state-driven (SD) correlation energy that will allow the use of cofe HEGs in density functional approximations.
As explained in the main text, we propose,
\begin{align}
E_{\xc}^{{\rm eLDA}}:=\int n(\vr) [\epsx(r_s)f_{\xrm}^{\cof}(\fb)
+ \epsc^{\SD,\cof}(r_s,\fb)]\;,
\label{eqn:Excot}
\end{align}
where $r_s(\vr)$ and $\fb(\vr)$ depend on local properties of the inhomogeneous system.

The exchange term involves the closed form
expression of eq.~\eqref{eqn:epsxcof}.
The correlation term, $\epsc^{\SD,\cof}(r_s,\fb)$, 
needs to be parametrised using:
\begin{enumerate}
\item The known high-density behaviour of eq.~\eqref{eqn:epschd};
\item The known low-density behaviour of eq.~\eqref{eqn:epscld}.
\item QMC data for other densities, adapted using eq.~\eqref{eqn:epscmhd};
\end{enumerate}
The high-density limit yields, [to $O(r_s\log(r_s))$]
\begin{align}
\epsc^{\SD,\cof}(r_s\to 0,\fb):=&c_0(\fb)\log r_s - c_1(\fb)
\end{align}
where the parameters $c_{0,1}(\fb)$ are linear in
$\fb$ and are trivially related to their un- and
fully-polarized counterparts.~\cite{PW92}
The low-density limit yields, [to $O(\tfrac{1}{r_s^2})$]
\begin{align}
\epsc^{\SD,\cof}(r_s\to \infty,\fb):=&
\tfrac{-C_{\infty}+C_{\xrm}[2/\fb]^{1/3}}{r_s}
 + \tfrac{C'_{\infty}}{r_s^{3/2}}\;.
\end{align}
where $C_{\infty}$, $C_{\xrm}$ and $C'_{\infty}$ are
universal parameters that do not depend on $\fb$.~\cite{Gould2023-ESCE}

Perdew and Wang~\cite{PW92} proposed that HEG
correlation energies lend themselves to a parameterization,
\begin{align}
F(r_s;P):=&-2A(1+\alpha r_s)\log\bigg[1+\frac{1}{2A\sum_{i=1}^4\beta_ir_s^{i/2}}\bigg]
\label{eqn:FPW92}
\end{align}
where $P=(A,\alpha,\beta_1,\beta_2,\beta_3,\beta_4)$ is a set
of parameters that depend on $\zeta$, $\fb$ or related variables.
By construction, eq.~\eqref{eqn:FPW92} can be made
exact to leading orders for small and large
$r_s$. The high-density limit yields,
\begin{align}
A^{\cof}=&c_0\;,
&
\beta^{\cof}_1=&\frac{e^{-c_1/(2c_0)}}{2c_0}\;,
&
\beta^{\cof}_2=&2A\beta_1^2\;,
\end{align}
where the coefficients are,
\begin{align*}
c_0(\fb)=&\tfrac{0.031091\fb}{2}\;,
&
c_1(\fb)=&0.00454 + \tfrac{0.0421\fb}{2}\;.
\end{align*}
The low-density limit yields,
\begin{align}
\beta^{\cof}_4=&\frac{\alpha}{C_{\infty}-C_{\xrm}f_{\xrm}^{\cof}(\fb)}\;,
&
\beta^{\cof}_3=&\frac{\beta_4^2C'_{\infty}}{\alpha}\;,
\end{align}
using the parameters
$C_{\infty}\approx 1.95 C_{\xrm}$ and
$C'_{\infty}=1.33$~\cite{Seidl2018}
from Sec.~\ref{sec:ldCorr}, and
$f_{\xrm}^{\cof}(\fb)=[2/\fb]^{1/3}$ from
eq.~\eqref{eqn:epsxcof}.
Thus, only $\alpha$ is left undefined.

\begin{table}[t!]
\caption{Correlation energy parameters for selected values of $\zeta$ from fits to benchmark data~\cite{Spink2013} and exact constraints.
\label{tab:Params}}
\begin{ruledtabular}\begin{tabular}{lrrrrrr}
$\zeta^{\QMC}$ & $A$ & $\alpha$ & $\beta_1$ & $\beta_2$ & $\beta_3$ & $\beta_4$ \\\hline
\multicolumn{7}{c}{cofe parameters} \\\hline
0.00 & 0.031091 & 0.1825 & 7.5961 & 3.5879 & 1.2666 & 0.4169 \\
0.34 & 0.028833 & 0.2249 & 8.1444 & 3.8250 & 1.6479 & 0.5279 \\
0.66 & 0.023303 & 0.2946 & 9.8903 & 4.5590 & 2.5564 & 0.7525 \\
1.00 & 0.015545 & 0.1260 & 14.1229 & 6.2011 & 1.6503 & 0.3954 \\
\hline
\multicolumn{7}{c}{rPW92 parameters} \\\hline
0.00 & 0.031091 & 0.1825 & 7.5961 & 3.5879 & 1.2666 & 0.4169 \\
0.34 & 0.030096 & 0.1842 & 7.9233 & 3.7787 & 1.3510 & 0.4326 \\
0.66 & 0.026817 & 0.1804 & 9.0910 & 4.4326 & 1.5671 & 0.4610 \\
1.00 & 0.015546 & 0.1259 & 14.1225 & 6.2009 & 1.6496 & 0.3952 \\
\end{tabular}\end{ruledtabular}
\end{table}

\begin{table}
\caption{Weighted sum parameters for $M_{2,3}$ (Appendix~\ref{app:Param}) and $Z_{2,3}$ (Appendix~\ref{app:rPW92}). E.g.,
$M_2=-2\epsc^0 + 4\epsc^{0.66}-2\epsc^1$ and $Z_3=19.86\epsc^0-30.57\epsc^{0.34}+12.71\epsc^{0.66}-2\epsc^1$.
\label{tab:Weights}
}
\begin{ruledtabular}\begin{tabular}{lrrrr}
Function & $\epsc^0$ & $\epsc^{0.34}$ & $\epsc^{0.66}$ & $\epsc^1$
\\\hline
\multicolumn{5}{c}{cofe parameters} \\\hline
$M_2$ & -2.00 & 0.00 & 4.00 & -2.00 \\
$M_3$ & 13.33 & -22.41 & 11.43 & -2.35 \\
\multicolumn{5}{c}{rPW92 parameters} \\\hline
$Z_2$ & -10.95 & 13.32 & -1.47 & -0.90 \\
$Z_3$ & 19.86 & -30.57 & 12.71 & -2.00 \\ 
\end{tabular}\end{ruledtabular}
\end{table}

Our goal is to find parameters, $P(\fb)$, that can
be used in a constant occupation factor (cof)
parameterization,
$\epsc^{\cof}(r_s,\fb^*):=F(r_s;P^{\cof}(\fb^*))$,
of the cofe HEG at selected values of $\fb^*$;
and interpolated to general $\fb$. Our first step
is to pick the values of $\fb^*$. 
We seek to
adapt the high-quality QMC data of
Spink~\emph{et al}~\cite{Spink2013},
who provided correlation energies for,
$\zeta^*\in(0, 0.34, 0.66, 1)$,
using eq.~\eqref{eqn:epscmhd}.
We therefore seek parametrizations at
$\fb^*=\fmap^{-1}(\zeta^*)$,
so that the right-hand side of
eq.~\eqref{eqn:epscmhd} is known.

As a first step, we must find $\fmap$
and its inverse. Setting
eqs.~\eqref{eqn:epsxzeta} and
\eqref{eqn:epsxcof} to be equal yields,
\begin{align}
\fmap(\zeta)\approx& 2-\tfrac43\zeta^2
+\tfrac16[1.0187\zeta^3+0.9813\zeta^4]\;,
\label{eqn:fmap}
\\
\fmap^{-1}(\fb)\approx& \sqrt{\tfrac{3}{4}(2-\fb)}
\big[ 1 + \big(\sqrt{\tfrac{4}{3}}-1\big)(2-\fb) \big]\;,
\label{eqn:fmapInv}
\end{align}
which are exact in the polarized and unpolarized
limits, and accurate to within 0.2\% for all
$\zeta$ and $\fb$.
Eq.~\eqref{eqn:fmap} gives
$\fb^*\in (2,1.85,1.50,1)$
for $\zeta^*\in(0, 0.34, 0.66, 1)$,
which are the $\fb$ values we use in fits.
Then, for each $\fb^*$, we obtain
$\alpha(\fb^*)$ by minimizing,
\begin{align}
\min_{\alpha}\sum_{r_s\in\text{QMC}}
\big|\epsilon_{\crm,\QMC}^{\zeta^{\QMC}}(r_s)
-\epsc^{\SD,\cof}(r_s,\fb^*)\big|
\end{align}
where $\epsilon_{\crm,\QMC}^{\zeta^{\QMC}}(r_s)$
is correlation energy data from
Ref.~\onlinecite{Spink2013}
and $\epsc^{\SD,\cof}(r_s,\fb^*)
:=F(r_s,P(\fb^*))$
involves the five constrained
coefficients and free $\alpha(\fb^*)$.
Optimal parameters for the four values
of $\zeta^*$ (called $\zeta^{\QMC}$ to
highlight their origin)
are reported in Table~\ref{tab:Params}.

The next step of our parametrisation departs from
PW92, in that we approximate the 
correlation energy at arbitrary $\fb$ via
cubic fits (in $\fb$) to the QMC data. Thus,
\begin{align}
&\epsc^{\cof}(r_s,\fb):=
(\fb-1)\epsc^{0}(r_s)
+ (2-\fb)\epsc^{1}(r_s)
\nonumber\\&
~~~~~~+ (\fb-1)(2-\fb)\big[M_2(r_s) + (\tfrac32-\fb)M_3(r_s)\big]\;,
\end{align}
where $\epsc^{\zeta}(r_s):=F(r_s,P_{\zeta})$ is computed using
eq.~\eqref{eqn:FPW92} and $M_2$ and $M_3$
involve weighted sums of $\epsc(r_s,\fb)$
at selected values of $\fb$.
This fit becomes exact in the high-density limit,
as the correlation energy is linear in $\fb$;
and is also extremely accurate in the low-density
limit as $(2/\fb)^{1/3}$ for $\fb\in[1,2]$ may
be reproduced to within 0.1\% by a cubic fit.
A cubic fit on $\fb^*\in (2,1.85,1.50,1)$ yields,
\begin{align}
M_2(r_s):=&2\big[2\epsc^{0.66}(r_s)-\epsc^0(r_s)-\epsc^1(r_s)\big]\;,
\\
M_3(r_s):=&\tfrac{40}{357}
\big[102\epsc^{0.66}(r_s)-200\epsc^{0.34}(r_s)
\nonumber\\&
+119\epsc^1(r_s)-21\epsc^0(r_s)\big]\;,
\end{align}
where $\alpha$ is optimized on each of the four
spin-polarizations.

The same strategy may also be applied to a
conventional spin-polarized HEG.
Thus, in addition to parameters for the cofe model,
Tables~\ref{tab:Params} and
\ref{tab:Weights} also contains a set of
coefficients for a ``revised PW92'' (rPW92)
model that is
an analogue of the cofe model introduced here.
Details are provided in Appendix~\ref{app:rPW92}.
As it is based on similar principles, rPW92 is
more directly comparable to the cofe parametrization
provided here than the original PW92,
especially in the low-density limit.

\section{Revised PW92}
\label{app:rPW92}

\begin{figure}
\includegraphics[width=\linewidth]{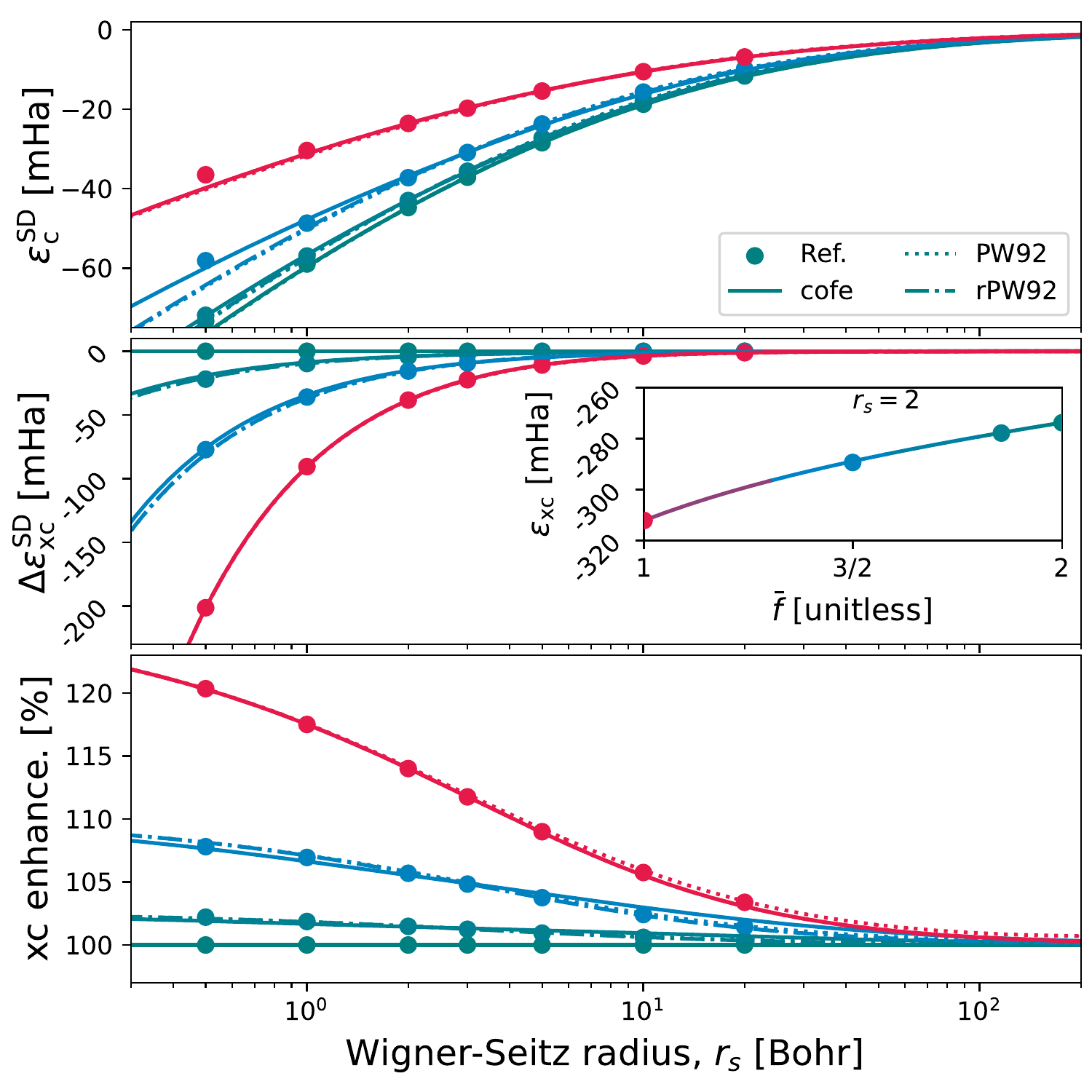} 
\caption{Like Figure~\ref{fig:epsxcParam} but with the addition of polarized HEG results from PW92~\cite{PW92} (dots) and rPW92 (dash-dot) for $\zeta\in(0,0.34,0.66,1)$, to show differences between cofe and polarized gases in the high-density limit.
\label{fig:epsxcParam_LSDA}}
\end{figure}

The ``revised PW92'' (rPW92) parameterization is designed as a direct replacement for the original PW92 model.~\cite{PW92}
Its main differences are:
1) the use of a cubic fit in $\zeta^2$, analogous to the fit to $\fb$ used in the main text;
2) the use of the most up-to-date understanding of the low density limit, per Sec.~\ref{sec:ldCorr};
and 3) $\alpha$ is found from the Spink reference data.~\cite{Spink2013}
Note, we fit to $\zeta^2$ because exchange and correlation are quadratic for $\zeta\to 0$, but linear for $\fb\to 2$.

The revised PW92 (rPW92) parameterization of correlation energies is,
\begin{align}
\epsc^{\rPW}(r_s,\zeta):=&(1-\zeta^2)\epsc^0+\zeta^2\epsc^1
\nonumber\\&
+(1-\zeta^2)\zeta^2\big[Z_2(r_s) + \zeta^2 Z_3(r_s)\big]\;.
\label{eqn:rPW92}
\end{align}
where coefficients for $Z_{2,3}$ are reported in Table~\ref{tab:Weights}.
Interestingly, the values we obtain for $\alpha$ at $\zeta=0$ and $\zeta=1$ are slightly lower than those from the original PW92 parametrisation,~\cite{PW92} most likely due to the use of more modern QMC data.

Figure~\ref{fig:epsxcParam_LSDA} shows results from Figure~\ref{fig:epsxcParam} plus the LSDA (rPW92) parametrised along similar lines.
It also includes results from an existing LSDA (PW92~\cite{PW92}).
By construction, both cofe and rPW92 do a better job of capturing the SCE limit, especially as PW92 incorrectly yields different low-density behaviours for different $\zeta$.
It is important to recognise that differences  (for $\zeta=0.34$ and $0.66$) between cofe enhancement factors and PW92/rPW92 \emph{do not represent errors}, but rather represent different quantum physics captured by cofe and polarized gases, which lead to different high density behaviours.

\nocite{Levi2020a}

\bibliography{HEG}

\end{document}